\documentclass{aa}
\usepackage[varg]{txfonts}

\usepackage[utf8]{inputenc}
\usepackage{hyperref}
\usepackage{textcomp}      
\usepackage{natbib}

\bibpunct{(}{)}{;}{a}{}{,} 
\providecommand{\abs}[1]{\lvert#1\rvert}

\def\Sat{{S}}
\def\A{\mathcal{A}}
\def\kms{\,km~s$^{-1}$}

\def\myr{$\,M_\odot$~yr$^{-1}$}
\def\Msun{\,M_\odot}
\def\Lsun{\,L_\odot}

\def\th{\mathrm{rel}}

\def\phantom{\textsc{phantom}}
\def\Tg{T_\mathrm{g}}
\def\Td{T_\mathrm{d}}

\begin{document}

\title{3D simulations of AGB stellar winds --- I. Steady winds and dust formation}

\author{ L.~Siess\inst{\ref{iaa}} \and W.~Homan\inst{\ref{iaa}} \and S.~Toupin\inst{\ref{iaa}} \and D.~J.~Price\inst{\ref{moca}}}
\institute{Institut d'Astronomie et d'Astrophysique, Universit\'e libre de Bruxelles (ULB), CP 226, 1050 Brussels, Belgium \label{iaa}
    \and School of Physics and Astronomy, Monash University, Vic. 3800, Australia \label{moca}}
\date{Received / Accepted}

\offprints{lionel.siess@ulb.be}

\abstract{}
{We present the implementation of the treatment of particle ejection and dust nucleation in the smoothed particle hydrodynamics (SPH) code \phantom{}. These developments represent the first step toward a more complete modeling of dust-driven winds emanating from asymptotic giant branch (AGB) stars that can be  used for comparison with high resolution imaging of these stars.}
{The AGB outflow is modeled by injecting the SPH particles from a spherical inner boundary. This boundary is a series of concentric shells, with the AGB star at its center, and the particles are positioned on these shells on the vertices of an isocahedron geodesic surface. The outermost shell is ejected with a predefined radial velocity, and subsequent lower shells replenish the ejected ones, all rotated randomly to improve the isotropy of the outflow. The physical properties of the particles on these shells are set by solving the 1D  analytic steady wind equations. The formation of dust is calculated starting from a compact chemical network for carbon-rich material, which creates the building blocks of the solid-state particles. Subsequently, the theory of the moments is used to obtain dust growth rates, without requiring knowledge on the grain size distribution.
}
{We tested our implementation against a series of 1D reference solutions. We demonstrate that our method is able to reproduce Parker-type wind solutions. For the trans-sonic solution, small oscillations are present in the vicinity of the sonic point, but these do not impact the trans-sonic passage or terminal wind velocity. Supersonic solutions always compare nicely with 1D analytic profiles. We also tested our implementation of dust using two formalisms: an analytic prescription for the opacity devised by Bowen and the full treatment of carbon-dust formation. Both simulations reproduce the 1D analytic solution displaying the expected additional acceleration when the gas temperature falls below the condensation temperature.}
{}

\keywords{stars: winds, outflows -- method: numerical -- hydrodynamics -- stars: AGB and post-AGB -- astrochemistry}

\authorrunning{Siess et al.}

\maketitle

\section{Introduction}

During the thermally pulsing asymptotic giant branch (AGB) phase, low- and intermediate-mass stars develop strong mass loss and eject their envelope to become a short-lived ($\la 10^4$yr) post-AGB star before ending their life as a white dwarf.

Stellar winds are key ingredients that dictate the evolution of the star but also contribute to the chemical evolution of the Galaxy by releasing the by-products of stellar nucleosynthesis into the interstellar medium (ISM).
These stellar winds manifest as nebulae surrounding their progenitor star. In binary systems the interaction of the wind of the evolved star with its companion leads to complex interactions, often resulting in a large-scale shaping of the nebula into specific morphologies such as spirals, discs, arcs or bipolar outflows \citep[e.g.][]{Mauron2006,Maercker2012,Kervala2016,Homan2020,Homan2021}.
These interactions can also affect the composition of the companion star as it potentially accretes a fraction of the wind material lost by the primary. It is commonly accepted that Carbon Enhanced Metal Poor (CEMP), CH, barium or yellow symbiotic stars to list only a few \citep[for a review of chemically peculiar low-mass binaries, see ][]{Jorissen2003} inherited their peculiar surface abundances from a the wind pollution from an AGB companion.

With the advent of high resolution imaging, we now have access to fine structural details of the circumstellar environments of AGB stars and their interpretation requires advanced modelling. The past years have seen a renewed interest in the study of the hydrodynamics of interacting winds in binary systems \citep[e.g.][]{Mohamed_Podsiadlowski_2012,AMUSE,deValBorro2017,Chen2018,ElMellah2020,MacLeod2020,Luis2020,Maes2021,Malfait2021,Aydi2022}. However, in these studies the dust formation process was not taken into account. To our knowledge, dust nucleation has only been included in a single 3D hydrodynamics code \citep{Mohamed_Podsiadlowski_2012}, and grain growth is now treated in the {\sf CO5BOLD} star-in-a-box models by \cite{Hofner2019}.

The increase in opacity associated with dust formation is a key ingredient to provide the required acceleration that carries the wind beyond the escape velocity of the AGB star. Detailed 1D models of dust-driven AGB winds have been developed over the past decades with increasing refinement \citep[e.g.][]{Bowen_1988,Fleischer_etal_1992,Hoefner_etal_1995,Andersen_etal2003,Hofner2003,Jeong2003,Hofner2008,Mattsson2010,Bladh2015,Bladh_atal_2019} and provided a clear picture of the phenomenon. It is now widely accepted that mass ejection proceeds in two steps.  Stellar surface pulsations actuated by large-scale convective motions of the stellar interior \citep{Freytag2017} provide the necessary kinetic energy to lift the stellar matter up to regions favourable for dust formation.
The opaque dust material absorbs the stellar photons and is radially accelerated outward. The collision of the grains with the surrounding gas drags it along, until it reaches the stellar escape velocity, resulting in a large-scale outflow freed from the gravitational attraction of the AGB star. For this scenario to be successful, condensation into the solid state should be fast enough to ensure that the gas does not fall back onto the star, requiring efficient cooling mechanisms to pull the condensation radius closer to the stellar surface.
This classical picture also assumes that gas and dust are strongly coupled which remains a valid assumption for high mass loss rates \citep[e.g.][]{Sandin2004}.

In this paper, we present the implementation of mass ejection by dust-driven stellar winds and the treatment of dust formation including equilibrium chemistry, into the smoothed particle hydrodynamics code \phantom{}. The chemical reactions governing molecule formation as well as the type of dust that forms strongly depend on the C to O abundance ratio (C/O).
This work focuses on carbon-rich mixtures (C/O > 1), whose chemistry is better known and has been extensively studied in the context of late-type stars \citep[see, e.\,g.,][]{Gail_etal_1984, Gail_Sedlmayr_1985, Gail_Sedlmayr_1987a, Gail_Sedlmayr_1988,  Dorfi_Hoefner_1991}.

This paper is the first in a series where a set of critical ingredients for the modelling of companion-perturbed dusty stellar winds will be implemented in the smoothed particle hydrodynamics code \phantom. The subsequent papers will revolve around the following topics: the treatment of gas-dust drift using the ``one-fluid approximation'' of hydrodynamics for small grains available in the code \citep{laibe2014a,Price2015}, the radiative transfer and the acceleration of the dust-component of the wind using by the attenuated stellar radiation field,
the incorporation of realistic cooling and heating rate of the gas and dust mixture, and a pulsating stellar atmosphere.

The current paper is organized as follows: in Sect.~\ref{sec:physics} we present the modelling of dust nucleation and dust growth and in Sect.~\ref{sec:hydro} the physics involved in the SPH hydrodynamics of dusty winds. The numerical treatment of particle ejection in \phantom{} is described Sect.~\ref{sec:numimp} and the validation of our approach is presented in Sect.~\ref{sec:3D} by comparing our 3D simulations with analytic profiles under various physical conditions and setups. We then conclude in Sect.~\ref{sec:summary} and discuss future prospects.

\section{Dust creation}
\label{sec:physics}

The physical ingredients and methods used to simulate dust-driven winds closely follows the description presented in the reference textbook of \citet{Gail_Sedlmayr_2014}. We consider a single-fluid approach where the grains are at rest with the gas particles and focus on mass loss from carbon-rich stars (C/O$>$1). For a consistent calculation of the fluid dynamics, the abundance and opacity of the dust grains has to be determined at every point in space. The chemistry of the wind and the calculation of the dust properties are presented in Sects.~\ref{sec:molabund} and \ref{sec:moment}. Knowledge of these quantities will then allow us to calculate the radiative acceleration as described in Sect.~\ref{sec:hydro}.

\subsection{Grain growth and destruction}
\label{gragro}

Classically, dust formation proceeds in two steps. First, the formation of critical clusters which serve as condensation seeds and then the growth of dust grains to macroscopic sizes by the addition of small entities called monomers. The nucleation process accounts for the formation of the seeds and involves species in chemical equilibrium. Such an equilibrium prevails as long as the clusters are small that is if their size $N \le N_0$. Grain growth thus proceeds through the addition of small clusters of size $N\le N_0$. In our calculations of C-based dust, the growth of graphite is assumed to proceed homogeneously by the addition of clusters of the same type (C$_i$-molecules where we consider $i=1,2$)
\begin{equation}
 \mathrm{C}_\mathrm{N} + \mathrm{C}_i  \rightarrow  \mathrm{C}_{\mathrm{N}+i},
\label{reac:C}
\end{equation}
and by the reactions involving C-based molecules of a different type (heteromolecular growth)
\begin{eqnarray}
\label{reac:C2H}
\mathrm{C}_\mathrm{N} + \mathrm{C}_2\mathrm{H}_2 & \rightarrow &  \mathrm{C}_{\mathrm{N}+2}  + \mathrm{H}_2,  \\
\mathrm{C}_\mathrm{N} + \mathrm{C}_2\mathrm{H}\    & \rightarrow &  \mathrm{C}_{\mathrm{N}+2}  + \mathrm{H},
\label{reac:C2H2}
\end{eqnarray}
This set of reactions corresponds to $N_0=4$.  In the following presentation of the theory, for the calculation of the density of dust grains, it is necessary to define the minimum size $N_l$ of a dust grains. 
This quantity is not easy to determine and following \cite{Gail_Sedlmayr_1988} we will use $N_l=1000$.

For a growth process to be energetically favourable, the cluster size must exceed a critical value $N_*$. The equilibrium abundance of clusters smaller than $N_*$  decreases with increasing $N$, they are thus unstable and do not participate to the dust production. Conversely, once the seed has reached the critical size, growth can proceed. It results that the cluster with size $N_*$ is the least abundant and because it has the slowest reaction rate, it will set the nucleation rate. The existence of this critical size is the reason why grain formation is a two step process. The formation of critical clusters is considered as a stationary process while the growth process has to be treated as a time-dependent problem.

The formation of C-based dust grains depends on the amount of atomic carbon and carbon-containing molecules available in the gas phase. We present the method for calculating these atomic and molecular abundances in the next section. The grain formation and growth is treated with the method of moments and described  in Sect.~\ref{sec:moment}.

\subsection{Molecular abundances}
\label{sec:molabund}

Apart from atomic carbon, only the molecules C$_2$, C$_2$H, and C$_2$H$_2$ were considered to contribute to the formation of carbonaceous dust \citep{Gail_etal_1984}. However, the formation of these molecules competes with the formation of other molecules containing carbon, e.\,g., HCN and CH$_4$ \citep[][ch.~10.4]{Gail_Sedlmayr_2014}, whose formation in turn competes with the formation of all molecules containing H, C, and N. Thus, for a consistent determination of the abundances of molecules contributing to the formation of C-based dust, a network of chemical reactions must be considered. In our code, we take into account 26 molecules H$_2$, OH, H$_2$O, CO, CO$_2$, CH$_4$, C$_2$H, C$_2$H$_2$, N$_2$, NH$_3$, CN, HCN, Si$_2$, Si$_3$, SiO, Si$_2$C, SiH$_4$, S$_2$, HS, H$_2$S, SiS, SiH, TiO, TiS, TiO$_2$ and TiS and 26 chemical reactions accounting for their formation (details about the chemical network is provided in Appendix~\ref{sec:chemistry}).
We assume that all these reactions are in chemical equilibrium. For example, to calculate the abundance of H$_2$, the law of mass action of the reaction $2\,\mathrm{H} \rightarrow \mathrm{H}_2$ in chemical equilibrium is used, which reads
\begin{equation}
    \label{pparth2}
    P_{\mathrm{H}_2} = P_\mathrm{H}^2 K_{\mathrm{H}_2},
\end{equation}
\citep[cf.\ Eq.~(10.16) of][]{Gail_Sedlmayr_2014}, where $P_{\mathrm{H}_2}$ and
$P_\mathrm{H}$ are the partial pressures of H$_2$ and H, respectively and
$K_{\mathrm{H}_2}$ is the dissociation constant of the reaction. The latter is given by
\begin{equation}
    \label{dissh2}
    K_{\mathrm{H}_2} = \mathrm{e}^{-\Delta G_{\mathrm{H}_2} / (\mathcal{R} \Tg)},
\end{equation}
where $\mathcal{R}$ is the universal gas constant, $\Tg$ the gas temperature, and
$\Delta G_{\mathrm{H}_2}$ is the change in Gibbs energy of the reaction,
\begin{equation}
    \label{gibbsen}
    \Delta G_{\mathrm{H}_2} = \Delta G_\mathrm{f}(\mathrm{H}_2) - 2 \Delta G_\mathrm{f}(\mathrm{H}).
\end{equation}
We take the values of the Gibbs energies of formation $\Delta G_\mathrm{f}$ for all atoms and molecules from the JANAF tables \citep{Chase_etal_1986}.

By use of relations of the form of Eq.~(\ref{pparth2}), the partial pressures of the different molecules can be expressed as a function of the partial pressures of the atomic species, so that, e.\,g., the total number of hydrogen atoms per unit volume $n_{\langle\mathrm{H}\rangle}$, atomic and bound in molecules, can be expressed as
\begin{equation}
    \label{partah}
    n_{\langle\mathrm{H}\rangle} k \Tg = P_\mathrm{H} + 2 P_\mathrm{H}^2 K_{\mathrm{H}_2},
\end{equation}
assuming that H$_2$ is the dominant hydrogenous molecule. Similarly, the total number density of nitrogen atoms $n_{\langle\mathrm{N}\rangle} = \epsilon_\mathrm{N} n_{\langle\mathrm{H}\rangle}$, where $\epsilon_\mathrm{N}$ is the nitrogen abundance by number relative to hydrogen, reads
\begin{equation}
    \label{partan}
    \begin{split}
        n_{\langle\mathrm{N}\rangle} k \Tg =& P_\mathrm{N} + P_\mathrm{N} P_\mathrm{H}^3
        K_{\mathrm{NH}_3} + P_\mathrm{N} P_\mathrm{C} K_\mathrm{CN} + \\
                                      & P_\mathrm{N} P_\mathrm{C} P_\mathrm{H}
                                      K_\mathrm{HCN} + 2 P_\mathrm{N}^2 K_{\mathrm{N}_2},
    \end{split}
\end{equation}
where $K_{\mathrm{NH}_3}$, $K_\mathrm{HCN}$, and $K_{\mathrm{N}_2}$ are the dissociation constants of the reactions of formation of NH$_3$, HCN, and N$_2$, respectively. The constants are calculated by use of Eq.~(\ref{dissh2}), substituting $\Delta G_{\mathrm{H}_2}$ by the change in Gibbs energy of the respective reaction, which is calculated according to relations of the form of Eq.~(\ref{gibbsen}).

Equations~(\ref{partah}) and (\ref{partan}), along with the equations for the total number densities of the other considered elements, $n_{\langle\mathrm{C}\rangle}$,
$n_{\langle\mathrm{O}\rangle}$, etc. (see Table~\ref{tab:abundances} for the list of considered species), form a system of equations whose solution gives the atomic partial pressures $P_\mathrm{el}$ of the elements ``el''. These atomic partial pressures are then used to calculate the molecular partial pressures $P_\mathrm{mol}$ of the molecules ``mol'' or, equivalently, their number densities $n_\mathrm{mol} = P_\mathrm{mol} / (k \Tg)$, $k$ being the Boltzmann constant, by means of Eq.~(\ref{pparth2}) and similar equations for the other molecules. These equations are   equivalent to the mass conservation equations for the species involved in the nucleation phase.

\subsection{Moment equations}
\label{sec:moment}
For the simulation of grain growth and destruction, we use the moment equations described by \citet{Gauger_etal_1990} \citep[see also][]{Gail_Sedlmayr_1988, Fleischer_etal_1992,Gail_Sedlmayr_2014}. This method does not require the calculation of the grain size distribution $f(N,t)$ itself, but uses its moments. For spherical grains, the  moment of order $i$ of the grain size distribution reads
\begin{equation}
    \mathcal{K}_i = \sum_{N = N_l}^\infty  N^{i / 3} f(N, t),
\end{equation}
where $N$ is the grain size expressing the number of monomers in a grain, $f(N, t)$ the number density of grains of size $N$ and $N_l\sim 1000$ the limit below which the cluster is not considered as a dust grain. The surface area of a grain composed of $N$ monomers can be estimated as
\begin{equation}
    \A_N = 4 \pi a_0^2\,N^{2 / 3}\ ,
\end{equation}
 where $a_0$ is the radius of a monomer inside a grain, which is derived from the density \textbf{$\rho_\mathrm{carb}$} of the grain material, i.\,e.
\begin{equation}
a_0 = [3 A_\mathrm{carb} m_\mathrm{u} / (4 \pi \rho_\mathrm{carb})]^{1 / 3},
\label{eq:a0}
\end{equation}
where $A_\mathrm{carb} = 12.011$ denotes the atomic weight of carbon in our case. For a density $\rho_\mathrm{carb} = 2.25\ \mathrm{g\,cm}^{-3}$ \citep[Smithsonian Physical Tables,][]{Forsythe_2003} we find $a_0 = 1.28 \times 10^{-8}\ \mathrm{cm}$. Global dust properties can be estimated from the knowledge of the moments. In particular,
\begin{itemize}
\item[a)] the mean grain density $n_d =\mathcal{K}_0$, that is the density of all dust grains with size $N \ge N_l$.
\item[b)] the average grain radius $\langle a\rangle = a_0 \mathcal{K}_1/\mathcal{K}_0$
\item[c)] the average grain surface $\langle \A_N\rangle= 4 \pi a_0^2 \mathcal{K}_2/\mathcal{K}_0$
\item[d)] the average particle size $\langle N\rangle=\mathcal{K}_3/\mathcal{K}_0$
\item[e)] the number density of monomers of size $N \ge N_l$ condensed into grains  $n_\mathrm{cond} =\mathcal{K}_3$.
\end{itemize}
As demonstrated in \cite{Gauger_etal_1990}, an elegant way of writing the moment equations is to normalize the moments by $n_{\langle H \rangle}$, the number of H atoms per unit volume
\begin{equation}
    \widehat{\mathcal{K}}_i = \frac{\mathcal{K}_i}{n_{\langle H \rangle}}= \frac{\mathcal{K}_i \, \bar{m}_\mathrm{H}}{\rho},
\end{equation}
where $\rho$ is the density and $\bar{m}_\mathrm{H}=  m_\mathrm{H} \sum_i A_i \epsilon_i$  the mean mass per H-atom calculated using the abundance by number relative to hydrogen $\epsilon_i$, atomic weight $A_i$ of species $i$ (see Table \ref{tab:abundances} for a list of the selected species) and $m_\mathrm{H}$ the mass of an hydrogen atom. An important consequence of the previous relation is that at each time, the abundance of free carbon atoms (relative to H) is known and given by
\begin{equation}
    \epsilon_\mathrm{C} = \epsilon_\mathrm{C}^0-\widehat{\mathcal{K}}_3,
    \label{eq:epsC}
\end{equation}
where $\epsilon_\mathrm{C}^0$ is the initial abundance at the beginning of the nucleation. With these normalized variables, the  evolution of the first four moments reads
\begin{eqnarray}
    \label{jevol}
    \frac{\mathrm{d}\widehat{J}_*}{\mathrm{d}t} & = &\frac{\widehat{J}_*^s - \widehat{J}_*}{\tau_*}, \\
    \label{moment0}
    \frac{\mathrm{d}\widehat{\mathcal{K}}_0}{\mathrm{d}t} & = &\widehat{J}_*, \\
    \label{momenti}
    \frac{\mathrm{d}\widehat{\mathcal{K}}_i}{\mathrm{d}t} & = & \frac{i\,  \widehat{\mathcal{K}}_{i-1}}{3 \tau} +N_l^{i/3}\widehat{J}_*,
\end{eqnarray}
where $\mathrm{d} / \mathrm{d}t$ is the material derivative. $\tau^{-1}$ is the net rate of growth or evaporation of the grains (see next section), $\widehat{J}_*$ is the normalized rate of formation of critical clusters per unit volume, $\widehat{J}_*^\mathrm{s}$ the normalized quasi-stationary rate, and $\tau_*$ the  relaxation time towards equilibrium (see sect.~\ref{nucrate}). The last term in Eq.~(\ref{momenti}) describes the transition from the (microscopic) newly formed clusters to the (macroscopic) grains with size $N>N_l$. As shown in \cite{Hofner_Dorfi_1992}, the inclusion of this term has a small impact. The solutions of Eqs.~(\ref{jevol}--\ref{momenti}) are computed analytically assuming that during the integration time step $\Delta t$, the quasi-stationary rate $\widehat{J}_*^\mathrm{s}$  is constant. We stress that  $\Delta t$ is constrained so this assumption holds.

\subsubsection{Net rate of grain growth and destruction}
\label{netrate}
For our C-rich chemistry, as described by \cite{Gauger_etal_1990}, the net rate of grain growth and destruction can be expressed as
\begin{equation}
\begin{split}
\frac{1}{\tau}  = & \  \frac{\A_1}{\sqrt{\pi}} \Biggl\{\alpha_1 \frac{P_\mathrm{C}}{m_\mathrm{C}\, v_{\th,\mathrm{C}}} \left(1 - \frac{1}{S}\frac{\alpha_*(\mathrm{C})}{b_\mathrm{C}} \right) \\
& + 2 \alpha_2\frac{P_{\mathrm{C}_2}}{m_{\mathrm{C}_2}\, v_{\th,\mathrm{C}_2}} \left(1-\frac{1}{S^2} \frac{\alpha_*(\mathrm{C}_2)}{b_{\mathrm{C}_2}}\right) \\
   & + 2 \alpha^c_{2,1} \frac{P_{\mathrm{C}_2\mathrm{H}}}{m_{\mathrm{C}_2\mathrm{H}} \, v_{\th,\mathrm{C}_2\mathrm{H}}}\left(1-\frac{1}{S^2}\frac{\alpha^c_*(\mathrm{C}_2\mathrm{H})}{b^c_{\mathrm{C}_2\mathrm{H}}} \right) \\
  & +2 \alpha^c_{2,2} \frac{P_{\mathrm{C}_2\mathrm{H}_2}}{ m_{\mathrm{C}_2\mathrm{H}_2}\, v_{\th,\mathrm{C}_2\mathrm{H}_2}} \left(1-\frac{1}{S^2}
           \frac{\alpha^c_*(\mathrm{C}_2\mathrm{H}_2)}{b^c_{\mathrm{C}_2\mathrm{H}_2}} \right) \Biggr\} ,
\end{split}
\label{taurate}\hspace{-1.3cm}
\end{equation}
where we only consider the chemical reactions given by Eqs.~(\ref{reac:C}), (\ref{reac:C2H}) and (\ref{reac:C2H2}) involving C, C$_2$, C$_2$H and C$_2$H$_2$.
The average relative velocity $v_{\th,i}$ between the reacting species $i$ (mass $m_i$) and the dust particle, in the absence of drift, is given by
\begin{equation}
    v_{\th,i} = \sqrt{\frac{2 k\Tg}{m_i}}\ .
    \label{eq_vth}
\end{equation}
The quantities $\alpha_1 = 0.37$ and $\alpha_2 = 0.34$ are the sticking coefficients of C and C$_2$ \citep{Thorn_Winslow_1957} and $\alpha^c_{2,i=1,2}$ the reaction efficiency of the considered molecules (C$_2$H or C$_2$H$_2$) with carbon. Following \cite{Gail_Sedlmayr_1988}, we set $\alpha^c_{2,1}=\alpha^c_{2,2}=\alpha_2$. The coefficients $\alpha_*(i)$ quantify the deviation from a thermal equilibrium population of states and are given by $\alpha_*(i)  = (\Td/\Tg)^{1/2}$, $\alpha^{c}_* = 1$ where $\Td$ and $\Tg$ refer to the dust and gas temperatures, respectively. The quantity $b_i$ accounts for departures of the particle densities from chemical and thermodynamical equilibrium between the solid (dust) and gas phase. Assuming chemical equilibrium, the coefficients $b_i$ are given by
\begin{align}
    b_\mathrm{C}& = \frac{\Td}{\Tg}, \\
    b_{\mathrm{C}_2}& =
        \frac{K_{\mathrm{C}_2}(\Tg)}{K_{\mathrm{C}_2}(\Td)}\frac{\Td}{\Tg}, \\
    b_{\mathrm{C}_2\mathrm{H}}^{c}& =
        \frac{K_{\mathrm{C}_2\mathrm{H}}(\Tg)}
        {K_{\mathrm{C}_2\mathrm{H}}(\Td)}, \\
    b_{\mathrm{C}_2\mathrm{H}_2}^{c}& =
        \frac{K_{\mathrm{C}_2\mathrm{H}_2}(\Tg)
        K_{\mathrm{H}_2}(\Td)}
        {K_{\mathrm{C}_2\mathrm{H}_2}(\Td)
        K_{\mathrm{H}_2}(\Tg)}.
\end{align}
The supersaturation ratio $\Sat$ is defined as the ratio of the partial pressure of carbon in the gas phase divided by the vapour pressure of solid carbon
\begin{equation}
    \Sat = P_\mathrm{C}(\Tg) / P^\mathrm{v}_\mathrm{C,solid}(\Td),
\end{equation}
where the saturation vapour pressure of carbon in the solid phase
\begin{equation}
    P^\mathrm{v}_\mathrm{C,solid} = K_\mathrm{\mathrm{C,solid}}^{-1} = \mathrm{e}^{\Delta G_\mathrm{C,solid} / (\mathcal{R} T)},
    \label{eq:Psat}
\end{equation}
is derived using the dust temperature ($T=\Td$). The free enthalpies of formation are taken from the JANAF tables and according to Eq.~(\ref{gibbsen}) reads $\Delta G_\mathrm{C,solid} = \Delta G_\mathrm{f}(\mathrm{C,solid}) - \Delta G_\mathrm{f}(\mathrm{C})$. See Appendix~\ref{detnet} for a more detailed explanation of Eq.~(\ref{taurate}).
When gas and dust are thermally coupled ($\Tg=\Td$), which is the assumption adopted in the current paper, the expressions for the growth rate simplifies since in this case $\alpha_*(i) = b_i = 1$. The full equations are however given so when radiative transfer is implemented, the expressions remain valid. Deviations from chemical and/or thermal equilibrium may be important in 3D simulations because of the development of structures (spirals, shocks, ...) that affect both the chemistry and  dust temperature and also because of the dependence of the grain growth rate on the gas-dust drift velocity \cite[see e.g.][]{kruger1997,Sandin2004}.

\subsubsection{Formation rate of seed particles}
\label{nucrate}

The quasi stationary rate of formation of critical clusters, i.\,e., clusters of size $N_*$, per unit volume reads \citep[chap 13.7.4 of][]{Gail_Sedlmayr_2014,Gail_etal_1984}
\begin{equation}
    \label{jstarnuc}
    J_*^\mathrm{s} = \beta \A_{N_*} Z\, \mathring{n}_\mathrm{d}(N_*).
\end{equation}
The quantity $\mathring{n}_\mathrm{d}(N)$ is the density of grains of size $N$ in thermal
equilibrium,
\begin{equation}
\mathring{n}_\mathrm{d}(N) =  n_\mathrm{C} \exp\left\{-\frac{\Delta F}{k\Tg}\right\}
\! =  n_\mathrm{C} \exp{\left\{(N\! -\! 1) \ln{\tilde{\Sat}}\! -\! \frac{\theta_N (N\! -\! 1)^{2 / 3}}{\Tg}\right\}},
\end{equation}
where $\Delta F$ is the free energy of formation of the grain from the vapour
\citep{Draine_Salpeter_1977}. The saturation ratio of the gas phase $\tilde{\Sat}$ is computed using $T=\Tg$ in Eq.~(\ref{eq:Psat}), $n_\mathrm{C}$ is the density of monomers, i.\,e., of C atoms, and the quantity $\theta_N$ is related to the surface tension $\sigma$ of the carbon grain material,
\begin{equation}
    \theta_N = \frac{\sigma\, \A_1 }{k\left\{1 + \left[N_h / (N - 1)\right]^{1 / 3}\right\}} = \frac{\theta_\infty}{1+(N_h/(N-1))^{1/3}},
\end{equation}
where we use a constant surface tension of $\sigma = 1400\ \mathrm{erg\,cm}^{-2}$ for carbon grains \citep{Tabak_etal_1975}. The surface of a monomer ($N=1$) is given by $\A_1$ and $N_h$ is the particle size for which the surface tension is reduced to one half of its value. Following \cite{Gail_etal_1984} (see also ch.~13.7 of \cite{Gail_Sedlmayr_2014} for a longer discussion), we set $N_h = 5$.

\begin{figure}
 \includegraphics{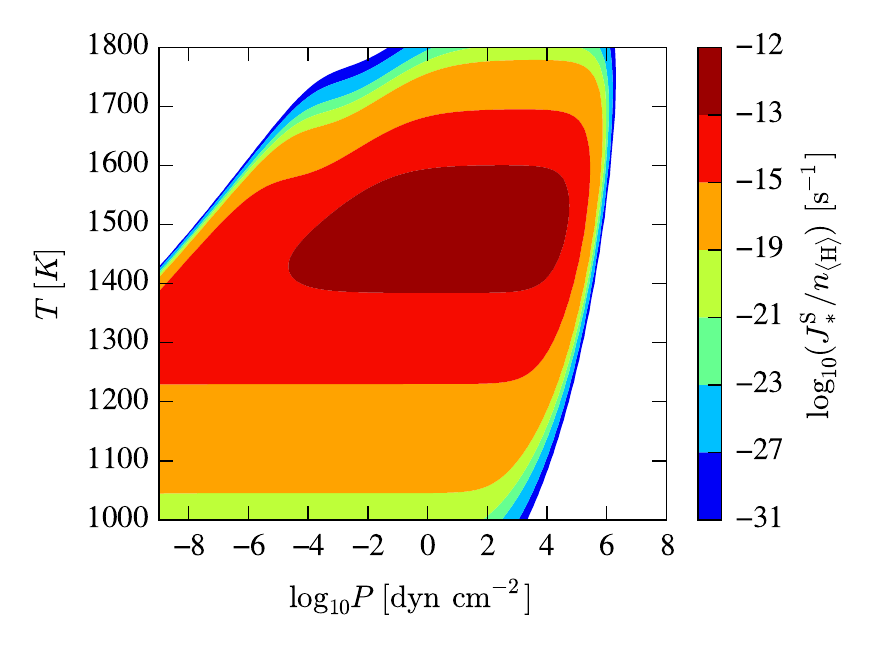}
 \caption{Nucleation rate per hydrogen atom ($J_*^\mathrm{s}/n_{\langle\mathrm{H}\rangle}$) as a function of temperature and pressure for a C/O ratio of 10. The computed values are nearly identical to the results presented in Fig.~5 of \citet{Gail_etal_1984}.}
    \label{nucleation}
\end{figure}

The size of the critical cluster $N_*$ is determined by the root of the equation $\partial \mathring{n}_\mathrm{d} / \partial N = 0 $. Its value is given by
\begin{equation}
    N_* = 1+\frac{N_{*,\infty}}{8}\left\{1+\left[1+2\left(\frac{N_h}{N_{*,\infty}}\right)^{1/3}\right]^{1/2}-2\left(\frac{N_h}{N_{*,\infty}}\right)^{1/3}\right\}^3,
\end{equation}
where
\begin{equation}
    N_{*,\infty} = (2\theta_\infty/3\Tg\ln \tilde{\Sat})^3\ .
\end{equation}
The Zeldovich factor $Z$ entering Eq.~(\ref{jstarnuc}) measures the deviation of the grain distribution from equilibrium and is defined by
\begin{equation}
    Z = \sqrt{\frac{1}{2 \pi} \left.\frac{\partial^2
        \ln{\mathring{n}_\mathrm{d}}}{\partial N^2}\right|_{N_*}}\
\end{equation}
and $\beta$, which gives the flux of monomers colliding with the grains, by
\begin{equation}
\beta = \frac{1}{\sqrt{2\pi k \Tg}} \left[\alpha_\mathrm{1} \frac{P_\mathrm{C}}{\sqrt{m_\mathrm{C}}} + 4 \alpha_2 \left(\frac{P_{\mathrm{C}_2}}{\sqrt{m_{\mathrm{C}_2}}} + \frac{P_{\mathrm{C}_2\mathrm{H}}}
            {\sqrt{m_{\mathrm{C}_2\mathrm{H}}}} + \frac{P_{\mathrm{C}_2\mathrm{H}_2}}
            {\sqrt{m_{\mathrm{C}_2\mathrm{H}_2}}}\right) \right]\ ,
\end{equation}
\citep{Gail_etal_1984}. With these quantities, the relaxation time towards equilibrium writes
\begin{equation}
    \tau_* = \left(2 \pi Z^2 \beta \A_{N_*}\right)^{-1}.
\end{equation}
As an illustration, we show in Fig.~\ref{nucleation} the nucleation rate calculated with Eq.~(\ref{jstarnuc}) as a function of temperature and pressure for a C/O ratio of 10. We use the chemical mixture of \citet{Ferrarotti_Gail_2002} which is very similar to that used by \citet{Gail_etal_1984}. The computed nucleation rates are nearly identical to the results presented in Fig.~5 of \citet{Gail_etal_1984}. Again, we find that nucleation mainly occurs around the carbon grain condensation temperature near 1500~K where the most abundant molecules are H$_2$, C$_2$H$_2$, and CO. With increasing temperature, the dust starts to be   dissociated and the nucleation rate drops. In the regime of low temperature on the other hand, dust formation is almost prohibited because the growth timescale becomes exceedingly long due to the very low density.

\begin{table}
  \caption{Elemental abundances by number relative to hydrogen  $\epsilon$ used in our
    models \citep{Ferrarotti_Gail_2002}.}
    \label{tab:abundances}
    \centering
    \begin{tabular}{l l}
        \hline\hline
        element & $\epsilon$ \\
        \hline
        He & $1.04 \times 10^{-1}$                     \\
        C  & $\epsilon_\mathrm{O} \times \mathrm{C/O}$ \\
        N  & $2.52 \times 10^{-4}$                     \\
        O  & $6.00 \times 10^{-4}$                     \\
        Si & $3.58 \times 10^{-5}$                     \\
        S  & $1.85 \times 10^{-5}$                     \\
        Ti & $8.60 \times 10^{-8}$                     \\
        Fe & $3.24 \times 10^{-5}$                     \\
        \hline
    \end{tabular}
\end{table}

\section{The physics of dust-driven winds}
\label{sec:hydro}

\subsection{3D hydrodynamics}
\label{3dhydro}

For our 3D hydrodynamic simulations we use the parallel smoothed particle hydrodynamics (SPH) code \phantom{} \citep{Price_Federrath_2010a, Lodato_Price_2010, Price2018}. In the framework of SPH, the hydrodynamic equations for a discretised fluid surrounding a central gravitational potential read \citep[see also the review by][]{Price_2012}
\begin{align}
    \label{density}
    \rho_i = & \sum_j m_j W(\abs{\vec{r}_i - \vec{r}_j}, h_i), \\
    \label{motion}
    \frac{\mathrm{D}\vec{v}_i}{\mathrm{D}t} = & - \sum_j m_j
        \left[\frac{P_i + q_i}{\Omega_i \rho_i^2} \nabla_i
        W_{ij}(\abs{\vec{r}_i - \vec{r}_j}, h_i)\right. \notag \\
    & \left. + \frac{P_j + q_j}{\Omega_j \rho_j^2} \nabla_i
        W_{ij}(\abs{\vec{r}_i - \vec{r}_j}, h_j)\right] - \frac{G M_*}{r_i^2}(1
        - \Gamma), \\
    \label{energy}
    \frac{\mathrm{D}e_i}{\mathrm{D}t} = & \frac{P_i}{\Omega_i \rho_i^2} \sum_j
        m_j (\vec{v}_i - \vec{v}_j) \cdot \nabla_i W_{ij}(\abs{\vec{r}_i - \vec{r}_j}, h_i) + \Lambda,
\end{align}
where the indices $i$ and $j$ denote an SPH particle and its neighbours, respectively. Each SPH particle carries the following properties: the local density $\rho_i$, its (constant) mass $m_i$, its position $\vec{r}_i$, smoothing length $h_i$, velocity $\vec{v}_i$, pressure $P_i$, artificial viscosity $q_i$, and specific internal energy $e_i$. $W(\abs{\vec{r}_i - \vec{r}_j}, h)$ is the smoothing kernel, $\nabla_i W_{ij}$ its gradient with respect to the location $\vec{r}_i$ of particle $i$, and $\Omega$ accounts for the gradient of the smoothing length.

The last term in Eq.~(\ref{energy}) represents the net cooling/heating rate of the SPH fluid. The last term in Eq.~(\ref{motion}) corresponds to the net acceleration due to the contributions from the gravitational attraction by the stellar mass $M_*$ with $G$ the gravitational constant, and from an outward force $\Gamma$, which is the ratio of the outward and gravitational accelerations. For a dust-driven stellar wind, the outward force exerted on the gas is generated by a two-stage mechanism. Firstly, the dust in the wind absorbs the incident stellar radiation, which deposits its momentum and propels the dust outwards. Secondly, the dust collides with the surrounding gas, and drags it along. In the optically thin limit, and assuming a full degree of mechanical coupling (also known as position coupling\footnote{In the position coupling regime, the dust and gas particles have the same position at all time and the gas-dust mixture is then one fluid. This also implies that gas and dust move at the same speed though the flow.}), it can be shown that
\begin{equation} \label{dustGamma}
    \Gamma = \frac{(\kappa_\mathrm{d} + \kappa_\mathrm{g}) L_*}{4 \pi c G M_*},
\end{equation}
where $\kappa_\mathrm{d}$ and $\kappa_\mathrm{g}$ are the Planck mean opacity coefficients of the dust (see following section) and gas, respectively, $L_*$ is the luminosity of the star, and $c$ the speed of light. Summing the gas and dust opacity contributions is consistent with our assumption of position coupling where the gas moves with the same speed as the dust. This remains a good approximation as long as the mass-loss rate is higher than $\sim 10^{-8}\ M_\sun\,\mathrm{yr}^{-1}$ \citep{Jorissen_Knapp_1998}.

The system of Eqs.~(\ref{density}) to (\ref{energy}) is supplemented with an equation of state (EOS) of the form $P = f(\rho, e)$. We assume an ideal gas equation, i.\,e.,
\begin{align}
    \label{ideal2}
    P&  = \frac{\rho}{\mu m_\mathrm{u}} k \Tg, \\
    \label{ideale}
    &  = (\gamma - 1) \rho e,
\end{align}
where $P$ is the total gas pressure, and $m_\mathrm{u}$ the atomic mass unit. For the calculation of the mean molecular weight $\mu$ and the polytropic index $\gamma$ of the gas it is sufficient to take into account only the most abundant species, i.\,e.\ H, He, and H$_2$, giving
\begin{equation}
    \mu = \frac{(1 + 4 \epsilon_\mathrm{He})\, n_{\langle \mathrm H\rangle} k \Tg}{P_\mathrm{H} + P_{\mathrm{H}_2} + \epsilon_\mathrm{He} n_{\langle \mathrm H\rangle} k \Tg},
    \label{eq:mu_def}
\end{equation}
and following \citet{Grassi2014}
\begin{equation}
    \gamma = \frac{5 P_\mathrm{H}+7 P_{\mathrm{H}_2}+5\epsilon_\mathrm{He}n_{\langle \mathrm H\rangle} k \Tg}{3 P_\mathrm{H}+5 P_{\mathrm{H}_2}+3\epsilon_\mathrm{He}n_{\langle \mathrm H\rangle} k \Tg}.
\end{equation}
In these expressions, $n_{\langle \mathrm H \rangle}$ is the total number of hydrogen atoms per unit volume in the medium and $\epsilon_\mathrm{He}$ is the total number of helium atoms per hydrogen atom. In this formula it is assumed  that H$_2$ is the dominant molecule carrying hydrogen, and that all helium remains monoatomic. The contribution from the other species is also neglected but this remains a good approximation given that they do not exceed 1\% of the total mass fraction. This also implies that the variations in $\mu$ and $\gamma$ will only result from the formation of molecular hydrogen. Below 500~K, we use a simplified chemical network where all hydrogen is in the form of H$_2$ and all carbon atoms locked in molecules (see Fig.~\ref{fig:abund}). This implies that dust formation is extremely inefficient at low temperatures yielding a very small value for the super-saturation ratio $S$.

We neglect the dust component in the EOS because the dust-to-gas ratio is at most 1\,\% in our simulations. Equation~(\ref{ideal2}) is exact when the interaction energy of the particles is small compared to their kinetic energy, which is the case in circumstellar dust shells \citep[ch.~3.4]{Gail_Sedlmayr_2014}.

In \phantom, the information carried by each SPH particles $i$ are by default $\vec{r}_i$, $\vec{v}_i$, $e_i$ and $h_i$ ($\rho_i$ is a simple function of $h_i$). To these variables, we add $J_{*,i}$, $\mathcal{K}_{0,i}$, $\mathcal{K}_{1,i}$, $\mathcal{K}_{2,i}$, $\mathcal{K}_{3,i}$, the mean molecular weight $\mu_i$, and the polytropic index $\gamma_i$. Because the chemistry couples the mean molecular weight of the gas to its temperature, and the temperature is obtained by assuming a certain value for $\mu_i$ (i.e. $\mu_i =\mu(T)$ and $T=T(\mu_i)$), our models initially suffered from numerical oscillations around the recombination temperature of H$_2$. These instabilities were eliminated after we introduced a sub-loop over $\mu_i$, $\gamma_i$ and $T$ to iteratively obtain continuous self-consistent values. Every other quantity (molecular abundances, dust fraction, opacity, etc.) can be recovered from these variables.

\subsection{Opacities}
\label{dustopac}

\subsubsection{Gas opacity}
\begin{figure}
    \includegraphics{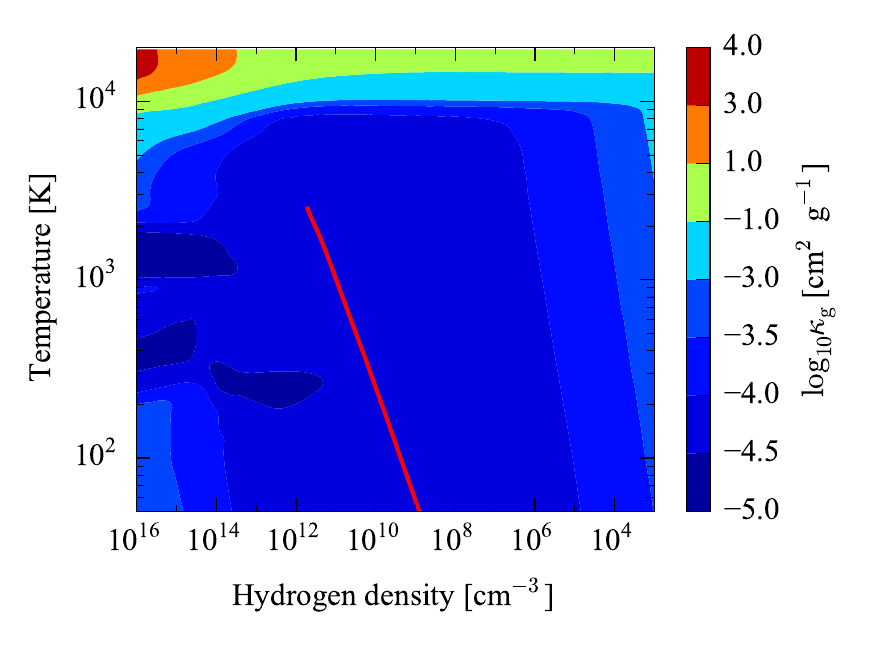}
    \caption{Planck mean values of the gas opacity $\kappa_\mathrm{g}$ (logarithmic scale), derived from a grid of models calculated with the Cloudy code using solar abundances with $\textrm{C/O = 1.4}$, and $J_\nu = B_\nu(3000~\mathrm{K})$. The thick red line traces a wind particle and illustrates the range of densities and temperatures encountered in our models.}
    \label{fig:kappag}
\end{figure}

 For the mean gas opacity $\kappa_\mathrm{g}$, we use the constant value proposed by \citet{Bowen_1988}, $\kappa_\mathrm{g} = 2 \times 10^{-4}\ \mathrm{cm}^2\,\mathrm{g}^{-1}$. To test the relevance of this approximation, we carried out detailed NLTE microphysics simulations over a wide range of densities and temperatures and confirm that the gas opacity is indeed of the order of $10^{-4}~\mathrm{cm^2\,g}^{-1}$ in our models. For these simulations, we used version 13.03 of the Cloudy code \citep{Ferland_etal_2013}, with a C abundance defined by $\textrm{C/O} = 1.4$, and solar abundances for the other elements. Figure~\ref{fig:kappag} shows a contour plot of the Planck mean gas opacity as a function of H number density and temperature under a mean intensity of $J_\nu = B_\nu (3000~\mathrm{K})$ representative of the radiation field of a central AGB star. The solid red line shows the trajectory of a wind particle wind in the $n_\mathrm{H}-T$ plane and demonstrates that the constant value of $\kappa_\mathrm{g}$ proposed by Bowen is a good approximation for our simulations. \cite{Winters_etal_2000} analyzed in detail the impact of molecular opacities and showed that using a lower opacity leads to higher mass loss rates and terminal velocities.

\subsubsection{Dust opacity}

In the framework of the Mie theory, the total opacity of a collection of spherical grains depends on the wavelength $\lambda$ and grain size $a$ and can be formulated as
\begin{equation}
\label{eq:Mie}
\kappa_\mathrm{d} = \frac{1}{\rho} \int_0^\infty \pi a^2 Q_\mathrm{ext}(a,\lambda) \, n(a)\, \mathrm{d}a\ ,
\end{equation}
where $n(a)$ is the grain size distribution function (cm$^{-4}$) and $Q_\mathrm{ext}$ the extinction efficiency. The determination of the opacity thus requires the knowledge of the distribution function. However, if the grain size is small compared to the wavelength ($\lambda$) of the stellar radiation field ($2\pi a/\lambda \ll 1$), in the so-called small particle limit (SPL), the extinction efficiency is dominated by absorption and in this Rayleigh regime it can be expressed as $Q_\mathrm{ext}(a,\lambda) = {Q}^\prime_\mathrm{ext}(\lambda)\,a$ allowing to separate in Eq.~(\ref{eq:Mie}) the wavelength and grain size dependence
leading to
\begin{equation}
\label{eq:kappad}
\kappa_\mathrm{d} = \frac{\pi}{\rho} Q^\prime_\mathrm{ext}(\lambda)\, \int_0^\infty a^3
n(a)\, \mathrm{d}a =  \frac{\pi a_0^3 }{\rho}\, Q^\prime_\mathrm{ext}(\lambda)\, \mathcal{K}_3,
\end{equation}
where $a_0$ is the radius of the monomer (Eq.~\ref{eq:a0}), and $\mathcal{K}_3$ is the third moment as described in Sect.~\ref{sec:moment}.
For an AGB star with an effective temperature of $\sim 2500-3000$~K, most of the flux is emitted around $1\mu$m. This implies that the SPL approximation will hold if $a \la 0.1\mu$m. Beyond this limit the scattering contribution must be included and the full Mie theory used. From their non-grey radiative transfer models, \cite{Hofner_Dorfi_1992} concluded that the SPL approximation remained valid. Later \cite{Mattsson_2011} mitigated this conclusion, mentioning that for models where the radiative-to-gravitational acceleration ratio is close to one, the SPL approximation breaks down but they also proposed a new approach that alleviates the limitation of the SPL approximation by using the Mie theory for one representative mean grain size. In future developments, we will implement this less restrictive and physically more consistent approach but if the grain size distribution shows a large variance, using one characteristic grain size may not give such a satisfactory result. Finally, it should be mentioned that recent models including non grey effects showed that scattering plays an essential role in launching  winds in (O-rich) M-type AGB stars \citep{Hofner2008,Bladh2015}, further emphasizing the need to go beyond the SPL approximation.

\begin{figure}
	\includegraphics[width=\columnwidth]{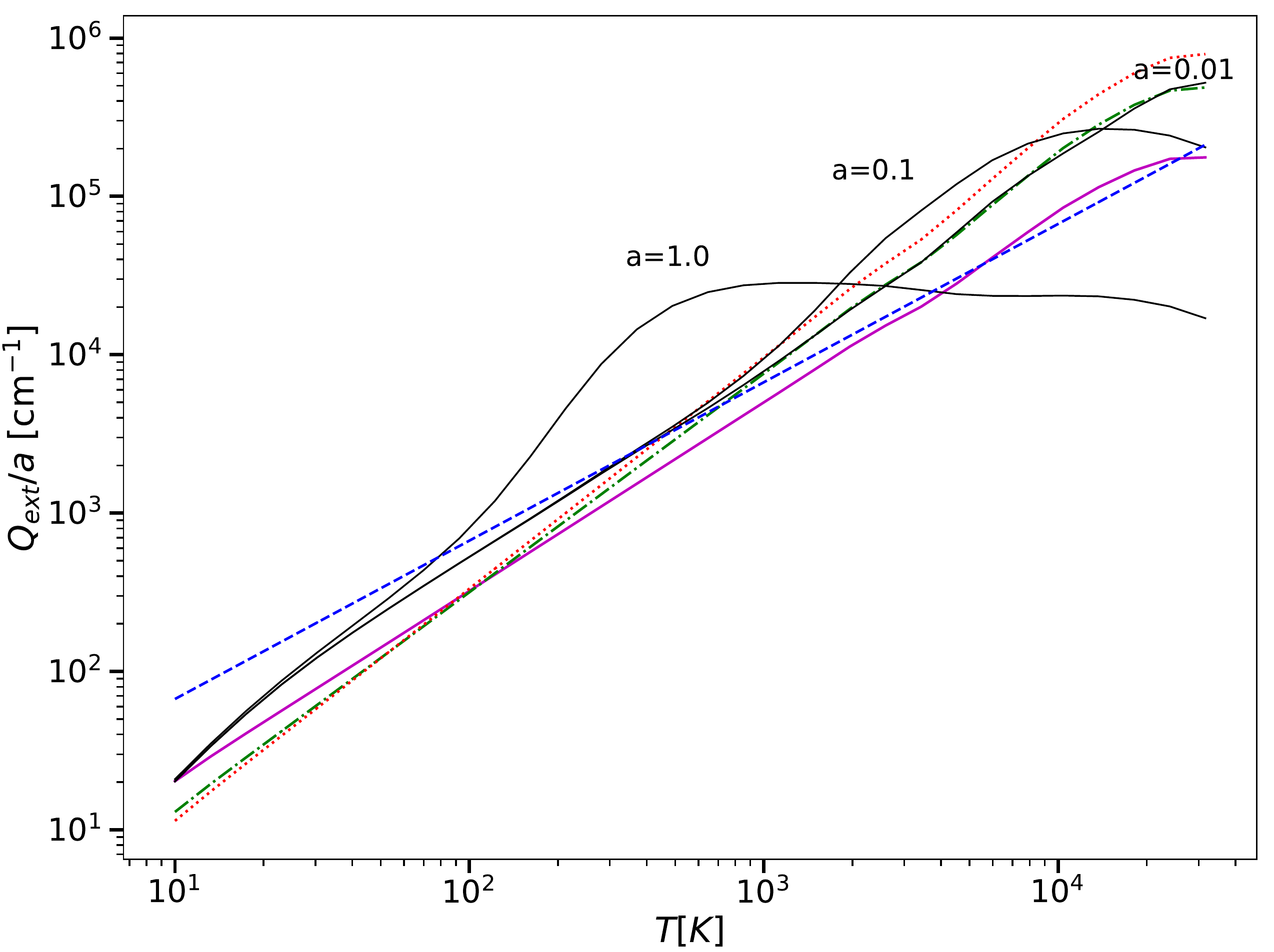}
    \caption{Planck mean values of the dust extinction efficiency $Q_\mathrm{ext}$ for amorphous carbon grains with size $a = 0.01$, 0.1 and $1$~{\textmu}m \citep[black solid lines, from][]{Preibisch_etal_1993}. For comparison, the approximations by \cite{Draine_1981} (blue) and \citet{Winters_etal_2000} with $n= 1.19$ (magenta), 1.38 (green) and 1.46 (red) are shown.}
    \label{fig:kappad}
\end{figure}

Various prescriptions are available for the Planck mean of these extinctions \citep{Gail_Sedlmayr_1987b} and in our simulations we use the fit given by \cite{Draine_1981}
\begin{equation}
 Q^\prime_\mathrm{ext}= 6.7\, \Td\ .
\end{equation}
Figure~\ref{fig:kappad} shows the Planck mean absorption efficiencies ($Q_\mathrm{ext}$) of amorphous carbon grains from various sources. The $Q_\mathrm{ext}$'s were calculated for three representative grains sizes ($a = 0.01$, 0.1 and $1 \mu$m) using the Mie theory for spherical grains (program BHMIE from \cite{Bohren_1983} adapted by Draine\footnote{\url{https://www.astro.princeton.edu/~draine/scattering.html}}). The optical data were taken from \cite{Preibisch_etal_1993}. These values are also compared to those obtained when approximating the wavelength dependence of the extinction efficiency in the SPL regime by a power law of the form  $Q^\prime_\mathrm{ext} =Q^\prime_0 (\lambda_0/\lambda)^n$ where the exponent varies between $n = 1.19,\ 1.38$ or 1.46 depending on the chosen optical data \citep{Winters_etal_2000}.
As can be seen, the quality of the fits deteriorates with increasing grain size, further illustrating the limitation of the SPL approximation. However, provided the grain size does not exceed $\sim 0.1 \mu$m,  the agreement between the different prescriptions is good.

\section{Numerical implementation in \phantom{} }
\label{sec:numimp}

\subsection{Inner wind boundary condition}
\label{sec:innerwind}

To simulate mass loss from a central source, particles are released in the numerical domain from a specified distance to the star's center called the injection radius $R_\mathrm{inj}$. The properties of the injected particles (density, velocity, internal energy and chemical composition) are set by solving the equations below for a radially expanding wind.

In this framework, the Lagrangian form of the hydrodynamic wind equations reads
\begin{align}
    \label{1d1}
    \frac{\mathrm{D}\rho}{\mathrm{D}t}& = - \frac{\rho}{r^2} \frac{\mathrm{d}(r^2 v)}{\mathrm{d}r}, \\
    \label{1d2}
    \frac{\mathrm{D}v}{\mathrm{D}t}& = - \frac{1}{\rho} \frac{\mathrm{d}P}{\mathrm{d}r} - \frac{G M_*}{r^2}(1 - \Gamma), \\
    \label{1d3}
    \frac{\mathrm{D}P}{\mathrm{D}t}& = - \frac{\rho c_\mathrm{s}^2}{r^2} \frac{\mathrm{d}(r^2 v)}{\mathrm{d}r} + (\gamma - 1) \rho \Lambda,
\end{align}
where $c_\mathrm{s}$ is the adiabatic sound speed defined by $c_\mathrm{s}^2 = \partial P / \partial \rho = \gamma P / \rho$.
Assuming a stationary wind, Eq.~(\ref{1d1}) reduces to the equation of mass conservation
\begin{equation}
    \label{dens1d}
    \rho(r) = \frac{\dot{M}}{4 \pi r^2 v},
\end{equation}
where $\dot{M}$ is the constant mass-loss rate. In this case, $\mathrm{D}P / \mathrm{D} t = v\,\mathrm{d}P / \mathrm{d}r$ and Eq.~(\ref{1d3}) can be used to replace the pressure gradient in Eq.~(\ref{1d2}), giving
\begin{equation}
    \label{momconcool}
    \frac{dv}{dr} = \frac{2 c_\mathrm{s}^2 / r - G M_* (1 - \Gamma) / r^2 - (\gamma - 1) \Lambda / v}{v (1 - c_\mathrm{s}^2 / v^2)}.
\end{equation}
The ideal gas law (Eq.~\ref{ideal2}) can be used to convert Eq.~(\ref{1d3}) into an equation for the temperature
\begin{equation}
    \label{tempcool}
    \frac{d\Tg}{dr} =(1 - \gamma) \Tg \left(\frac{2}{r} + \frac{1}{v}\frac{\mathrm{d}v}{\mathrm{d}r}\right) + \frac{(\gamma - 1)\, \mu m_\mathrm{u}}{k}\frac{\Lambda}{v}\ .
\end{equation}
Finally the time evolution is given by
\begin{equation}
    \label{radtime}
    \frac{dr}{dt} = v,
\end{equation}
We integrate Eqs.~(\ref{momconcool}), (\ref{tempcool}) and (\ref{radtime})
using the fourth-order Runge-Kutta method, to obtain the physical properties of the ejected SPH particles as a function of both distance and time.
For trans-sonic wind solutions, the numerator and denominator of Eq.~(\ref{1d2}) vanish at the sonic radius leading to an undefined fraction that causes numerical problems. To find the trans-sonic trajectory, the wind velocity at the inner boundary is varied until the numerator and the denominator are smaller than a threshold value of $10^{-2}$ in the vicinity of the sonic point. Then, L'H\^opital's rule is applied to obtain the correct velocity gradient.

A free wind is obtained when the velocity gradient given by Eq.~(\ref{momconcool}) is greater than or equal to zero beyond the sonic radius. However, there exist a set of solutions to Eqs.~(\ref{momconcool}), (\ref{tempcool}) and (\ref{radtime}) for which the velocity gradient is smaller than zero, which are known as breeze-type solutions \citep[see region 3 in Fig.~3.1]{Lamers_Cassinelli_1999}. These appear when the numerator and the denominator of Eq.~(\ref{momconcool}) have opposite signs. When $v<c_\mathrm{s}$, that is when the wind is subsonic, the breeze is obtained as long as the outward acceleration $\Gamma$ is higher than the threshold value
\begin{equation}
    \label{gammamax}
    \Gamma_\mathrm{max} = 1-\frac{4c_\mathrm{s}^2}{v_\mathrm{esc}^2}+(\gamma-1)\Lambda\frac{r^2}{GM_*v},
\end{equation}
where $v_\mathrm{esc}=(2GM_*(1-\Gamma)/R_*)^{1/2}$ is the escape velocity.

Our SPH simulations are unable to reproduce breeze-type solutions. Even if particles are launched from a hot corona with a velocity below the trans-sonic value, they ultimately converge on the trans-sonic solution. We speculate that the reason for this behavior is due to the fact that a breeze solution has a non vanishing pressure at infinity in contrast to the trans- and supersonic cases. Since our SPH simulations assume a free boundary condition and therefore zero pressure at infinity, the particles tend to converge toward that solution. A fixed outer boundary providing a non-zero outer pressure could be implemented using a method similar to the ``handled-shells'' method for particle injection described in the following section. However, a subsonic breeze was not detected in the solar system by the Mariner 2 space craft \citep[e.g.][]{Neugebauer1962} and this solution is also formally unstable \citep{Velli2001}. Given the limited interest of such types of wind, they are excluded from our study.

\subsection{Injection of SPH particles}
\label{inject}

At the inner boundary, the injected SPH particles are homogeneously distributed on a shell of radius $R_\mathrm{inj}$.
Taking a constant number $p$ of particles per shell, the time interval $\delta t$ between two subsequent shells is given by the relation
\begin{equation}
    \label{dtsphere}
    \delta t = p m / \dot{M},
\end{equation}
where $m$ is the constant mass of a SPH particle and $\dot{M}$ the wind mass loss rate. In this framework, the $k$'th shell is injected at time $\tilde{t}_k = k \delta t$. Hence, the number of shells injected at every time step is set by the fraction of $\delta t$ over the SPH timestep $\Delta t$.

Using the age of the simulation as a way to index the shells, we can track how consecutive shells should be launched. If $t^{n+1}$ is the time to be reached and $\Delta t^n$ the value of the current hydro (SPH) time step, then the shells with index $i_\mathrm{min}$ to $i_\mathrm{max}$ have to be ejected, where
\begin{align}
    \label{imin}
    i_\mathrm{min}& = \left\lfloor\frac{t^{n+1} - \Delta t^n}{\delta t}\right\rfloor + 1, \\
    \label{imax}
    i_\mathrm{max}& = \left\lfloor\frac{t^{n+1}}{\delta t}\right\rfloor,
\end{align}
(``$\lfloor\;\rfloor$'' is the floor function). When $\Delta t < \delta t$, $i_\mathrm{min}$ may become greater than $i_\mathrm{max}$, in which case no shell is released.

However, in general the hydrodynamical time $t^{n+1}$ does not coincide with the precise time $\tilde{t}_{i}$ at which injection should occur. Consequently, if $t^{n+1}>\tilde{t}_i$ the properties of the released particles have to be synchronized and estimated at the ``current'' time, that is at a time $t_e=t^{n+1}-\tilde{t}_i$ after their ejection. The velocity is given by the solution $v(t_e)$ of Eq.~(\ref{momconcool}), and the density by Eq.~(\ref{dens1d}), converted to $\rho(t_e)$ using the solution of Eq.~(\ref{radtime}). The internal energy (see Eqs.~(\ref{ideal2}) and (\ref{ideale})) is calculated using the temperature from Eq.~(\ref{tempcool}). The radius $r(t_e)$ at which the new shell is freed is given by Eq.~(\ref{radtime}).

\begin{figure}
    \includegraphics{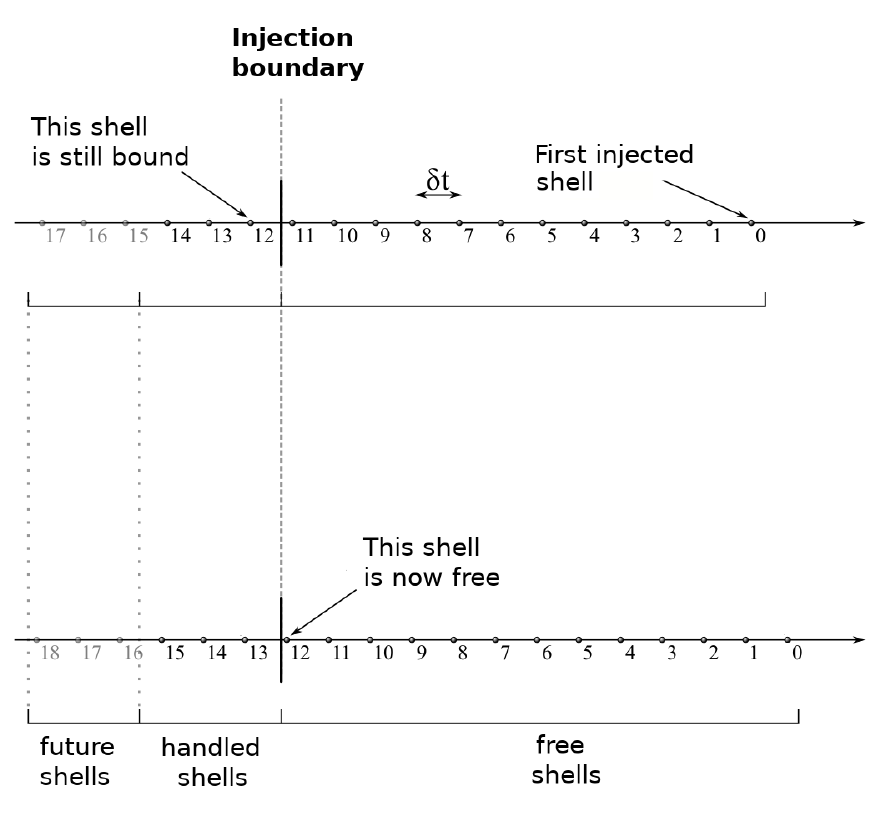}
    \caption{Schematic representation of the method for injecting SPH particles. The horizontal axes represent the time since ejection, or interchangeably the distance from the boundary. The numbered dots represent particle shells, numbered by their index. Older shells have a lower index. \emph{Top:} Before the hydro timestep the shell with index 12 lies below the injection boundary, and is therefore part of the handled shells, and has imposed physical properties. \emph{Bottom:} After the hydro timestep, the shell with index 12 has moved beyond the injection boundary, and no longer has imposed physical properties.}
    \label{fig:shells}
\end{figure}

In our SPH simulations, the central star is represented by a sink particle \citep{Bate1995,Price2018}, which can not exert pressure on its surroundings, and thus produces a central vacuum. This implies that any particle in the vicinity of the sink will feel a positive pressure gradient pulling material towards the sink. Consequently, simulating a wind by adding only one shell of SPH particles at a time will not result in an effective particle injection.
To compensate for this effect, a number of radially fixed, consecutive shells are placed around the sink. At each time step the density, velocity, and internal energy of the particles on these boundary shells are set to the analytical radial wind solution derived in Sect.~\ref{sec:innerwind}. This procedure ensures a correct behaviour of the physical variables at the inner boundary and in particular of the pressure gradient. The farthest handled shell dictates the radius beyond which the properties of the SPH particles are no longer imposed by the analytical solution and can evolve freely, as dictated by the SPH solver. This is the effective injection radius $R_\mathrm{inj}$. The number of fixed boundary shells can be specified in the code with an integer parameter ({\sf iboundary\_shells}). Our tests show show that 5 shells are able to accurately provide the expected pressure gradient.

In Figure~\ref{fig:shells}, we show a schematic diagram of what happens when the hydrodynamical time step $\Delta t$ is greater than the theoretical time between the release of consecutive shells $\delta t$. In this example three handled shells define the inner boundary, and shell with index number~12 is released.

\subsection{Distributing particles on geodesic spheres}

To distribute a finite number of particles as uniformly as possible on the surface of a sphere, we employ the icosahedron interpolation method \citep{Wang2011}. Using this method the triangular faces of an icosahedron are tiled with smaller triangles. The vertices of all these triangles are then projected onto the platonic shell of the original icosahedron, and the SPH particles are placed on the projected vertex coordinates.

Each triangular face of the icosahedron is subdivided in such a way that every edge contains $2 q$ vertices, including the corner points. Thus, there are $20 (2 q - 1)^2$ triangular faces and $p = 10 (2 q - 1)^2 + 2$ vertices in each shell\footnote{A comprehensive description of this tiling is provided (in french) at \texttt{https://fr.wikipedia.org/wiki/G\'eode\_géométrie}}.
To ensure that the projected vertices are as equidistant as possible we use the mapping procedure described by \citet{Tegmark_1996}, which ensures that every projected facet has the same area.
This is necessary to reduce initial inhomogeneities on the sphere and numerical artifacts in the simulations.

As the geodesic spheres are not perfectly isotropic, it is necessary to rotate each sphere with respect to the previous one to
prevent anisotropies from building up. The optimal rotation angles $\vartheta_i$ ($i = 1, 2, 3$) that will maximize the minimal distance between vertices of two subsequent geodesic spheres is given for various values of $q$ in Table \ref{rotang}.
We determined those angles empirically by testing thousands of random combinations.
\begin{equation}
    \hat{\vec{o}}' = \begin{pmatrix}
        c_2 c_3 & -c_2 s_2 & -s_2 \\
        -s_1 s_2 c_3 + c_1 s_3 & s_1 s_2 s_3 + c_1 c_3 & -s_1 c_2 \\
        c_1 s_2 c_3 + s_1 s_3 & -c_1 s_2 s_3 + s_1 c_3 & c_1 c_2
    \end{pmatrix} \hat{\vec{o}},
\end{equation}
where $c_i = \cos{\vartheta_i}$ and $s_i = \sin{\vartheta_i}$. $\hat{\vec{o}}$ is defined as a unit vector pointing towards an SPH particle on a sphere, and the subsequent sphere will be identical to the previous but rotated in such a way that the vector changes into $\hat{\vec{o}}'$.

\begin{table}
  \caption{Optimized rotation angles (in rad) for the distribution of SPH particles at the inner boundary as a function of the number of vertices or particles ($N_p=10(2q-1)^2+2$).}
    \label{rotang}
    \centering
    \begin{tabular}{l l l l l}
        \hline\hline
        $q$ & $N_p$ & $\vartheta_i$ & $\vartheta_j$ & $\vartheta_k$ \\
        \hline
        1   & 30   & 1.286936103  & 2.978630878  & 1.039528355  \\
        2   & 110  & 1.227187223  & 2.582394661  & 1.053604227  \\
        3   & 270  & 0.235711384  & 3.104772874  & 2.204402209  \\
        4   & 510  & 3.052314457  & 0.397072776  & 2.275006169  \\
        5   & 830  & 0.137429598  & 1.998606705  & 1.716093916  \\
        6   & 1230 & 2.904432935  & 1.779396863  & 1.041130506  \\
        10  & 3630 & 2.409130709  & 1.917210104  & 0.899557512  \\
        15  & 8430 & 1.956058284  & 0.110825899  & 1.911748564  \\
        \hline
    \end{tabular}
\end{table}

\subsection{Wind resolution} \label{ress}

There is only a finite number of ways in which particles can be arranged onto a shell using the icosahedron interpolation method described in the previous section. This inherent quantization implies that the tangential resolution of the inner boundary condition can only take on discrete configurations. It can be shown that the distance $d_\mathrm{\perp}$ between neighbouring particles on such a shell of unit radius is equal to
\begin{equation} \label{equi}
    d_\mathrm{\perp} = \frac{2}{(2 q - 1)} \left(\varphi \sqrt{5}\right)^{-1/2},
\end{equation}
where $\varphi = (1 + \sqrt{5}) / 2$ is the golden ratio, and $q$ can be any natural number, which represents the tangential resolution set-up, and appears in the \phantom{} parameter card as the {\sf iwind\_resolution} variable.

Conversely, there is no geometrical constraint on the resolution long the radial axis. Given the inherent length scale associated with Eq.~(\ref{equi}), it is natural to express the resolution along the radial axis as a function of this distance. In \phantom, the radial distance between two consecutive shells is defined as
\begin{equation} \label{shell}
    \delta r = w_\mathrm{ss} d_\mathrm{\perp}\ ,
\end{equation}
where $w_\mathrm{ss}$ can take on any value larger than zero, and can also be set in the \phantom{} parameter card though the variable {\sf wind\_shell\_spacing}.

Ideally, the resolution parameters should be chosen such that two primary requirements are satisfied. Firstly, the smoothing length of the injected SPH particles should be larger than the distance between them, both radially and tangentially. This is necessary to ensure that the particle kernels overlap and effectively feel each other along these directions. Secondly, the representative length-scale of the system must be resolved. We address how to approach both these requirements in wind systems.

The smoothing length $h$ is given by $h_\mathrm{fact}\, n^{-1/3}$ \citep{Price2018}, where $n$ is the particle number density, and $h_\mathrm{fact}$ is a proportionality factor that can vary with the adopted kernel and which is set by default to 1.2. As a consequence, $h$ can be re-written as
\begin{equation}
    h = h_\mathrm{fact}\, d_r^{\frac{1}{3}} d_{\perp}^{\frac{2}{3}},
\end{equation}
where $d_{\perp}$ is the tangential particle distance given by Eq.~(\ref{equi}), and $d_r \approx \delta r = w_{ss} d_\mathrm{\perp}$ is the radial distance between particle on adjacent shells. Hence, the first resolution requirement is satisfied when both the $h \geq d_r$ and $h \geq d_{\perp}$ conditions are simultaneously fulfilled. Because $d_r$ depends on $d_{\perp}$, this translates to a resolution condition on $w_\mathrm{ss}$ given by
\begin{equation} \label{wsscond}
    h_\mathrm{fact}^{-3} \leq w_{ss} \leq h_\mathrm{fact}^\frac{3}{2}.
\end{equation}

The second resolution requirement can be satisfied by choosing the parameter $q$ in Eq.~(\ref{equi}) such that the radial distance between consecutive shells is smaller than the representative length-scale of the system. For a Parker wind \citep{Lamers_Cassinelli_1999}, this length-scale can be identified to the sonic radius. For a wind-Roche-lobe-overflow binary \citep{Mohamed_Podsiadlowski_2012}, the representative length-scale would be set by the position of the first Lagrange point and for detached binaries by the orbital separation. Since a typical number of handled shells (see Sect.~\ref{inject}) of five is required to properly generate the inner boundary condition on the pressure, one should aim towards setting the parameter $q$ such that $\sim$ ten shells can fit between the inner boundary and the relevant critical radius.

Given the above constraints on the spatial resolutions, the mass of the SPH particles (which is typically considered as the standard measure for the resolution in SPH), is fully constrained by the choice of the mass-loss rate and initial wind velocity which are input parameters of \phantom. This is because the distance $\delta r$ between two adjacent shells can also be approximately given by
\begin{equation}
    \delta r \simeq v_\mathrm{inj}\,N_p m / \dot{M},
\end{equation}
where $v_\mathrm{inj}$ is the wind velocity at the injection radius $R_\mathrm{inj}$ (also available in the \phantom{} parameter card as {\sf wind\_velocity}), and $N_p$ is the number of particles\footnote{The number of particles on the isocahedron sphere is given by $N_p = 10(2q-1)^2+2$ where the parameter $q$ corresponds to the variable {\sf iwind\_resolution} in the \phantom{} input file.} of mass $m_p$. Given that the inter-shell distance is also given by Eq.~(\ref{shell}), this implies that the mass of the particles is determined by the set-up of the boundary conditions and
\begin{equation}
    m_p = w_\mathrm{ss} \frac{d_\mathrm{\perp}}{v_\mathrm{inj}}\frac{\dot{M}}{N_p}\ .
\end{equation}
The user can alternatively impose the particle mass and the code will find the closest value of the parameter $q$ that fulfills the above resolution criteria.

\begin{table}
    \caption{1D model parameters. $T_\mathrm{iso}$ corresponds to the temperature of the isothermal wind. }
    \label{modpar1d}
    \centering
    \begin{tabular}{l l}
        \hline\hline
        Parameter & Value \\
        \hline
        $\dot{M}$         & $10^{-5} M_\sun\,\mathrm{yr}^{-1}$ \\
        $M_*$             & $1.2\ M_\sun$           \\
        $L_*$             & $2 \times 10^4\ L_\sun$ \\
        $R_*$             & 1~au                    \\
        $R_\mathrm{inj}$  & 2~au                    \\
        $T_\mathrm{iso}$  & $1.25 \times10^{4}$~K      \\
        $\mu$             & 0.6$^\dag$ \\
        Kernel            & quintic              \\
        \hline
    \end{tabular}

    $^\dag$ This value of the mean molecular weight corresponds to a solar mixture where all the H is ionized.
\end{table}

\section{Benchmarking}
\label{sec:3D}

To validate the implementation of the above-mentioned methods, we calculate a sample of benchmarking simulations and compare the analytic solution to the 3D SPH wind profiles. In all simulations shown here, the net cooling rate $\Lambda$ in Eqs.~(\ref{energy}), (\ref{1d3}) and (\ref{momconcool}) is set to zero. Furthermore, the quintic kernel is used; its larger compact support radius
contributes to reduce the dispersion of the SPH solution.

\subsection{Stationary dust-free winds}

\begin{figure}
    \centering
    \includegraphics[width=9.5cm]{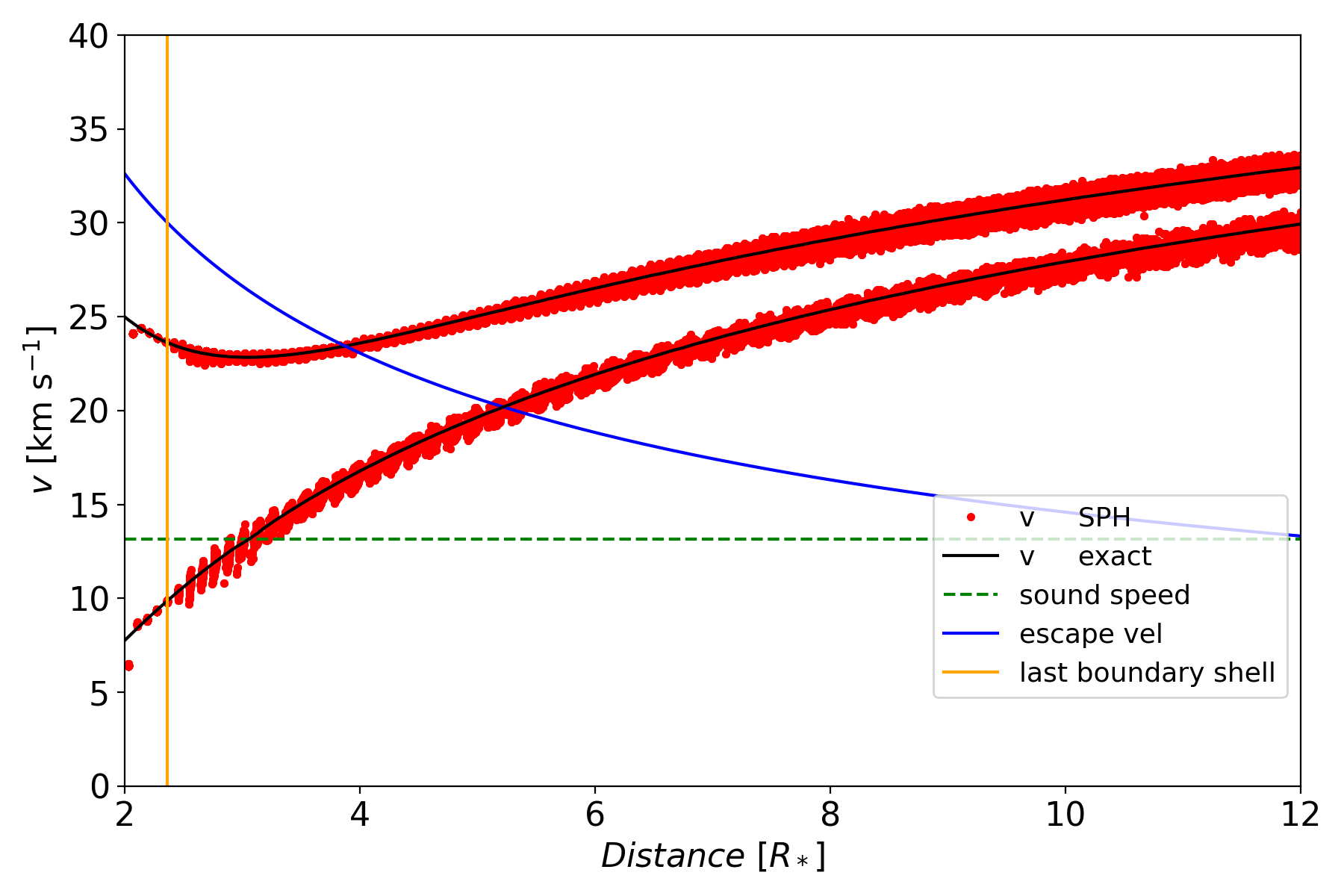}
    \caption{Isothermal velocity profiles of a transonic (\emph{bottom}) and supersonic (\emph{top}) SPH wind solution, in red. The analytic solution is given by the black curves for a wind temperature $T_\mathrm{iso} = 1.25\times 10^4$~K.}
    \label{fig:iso}
\end{figure}

\begin{figure}
    \centering
    \includegraphics[width=9.5cm]{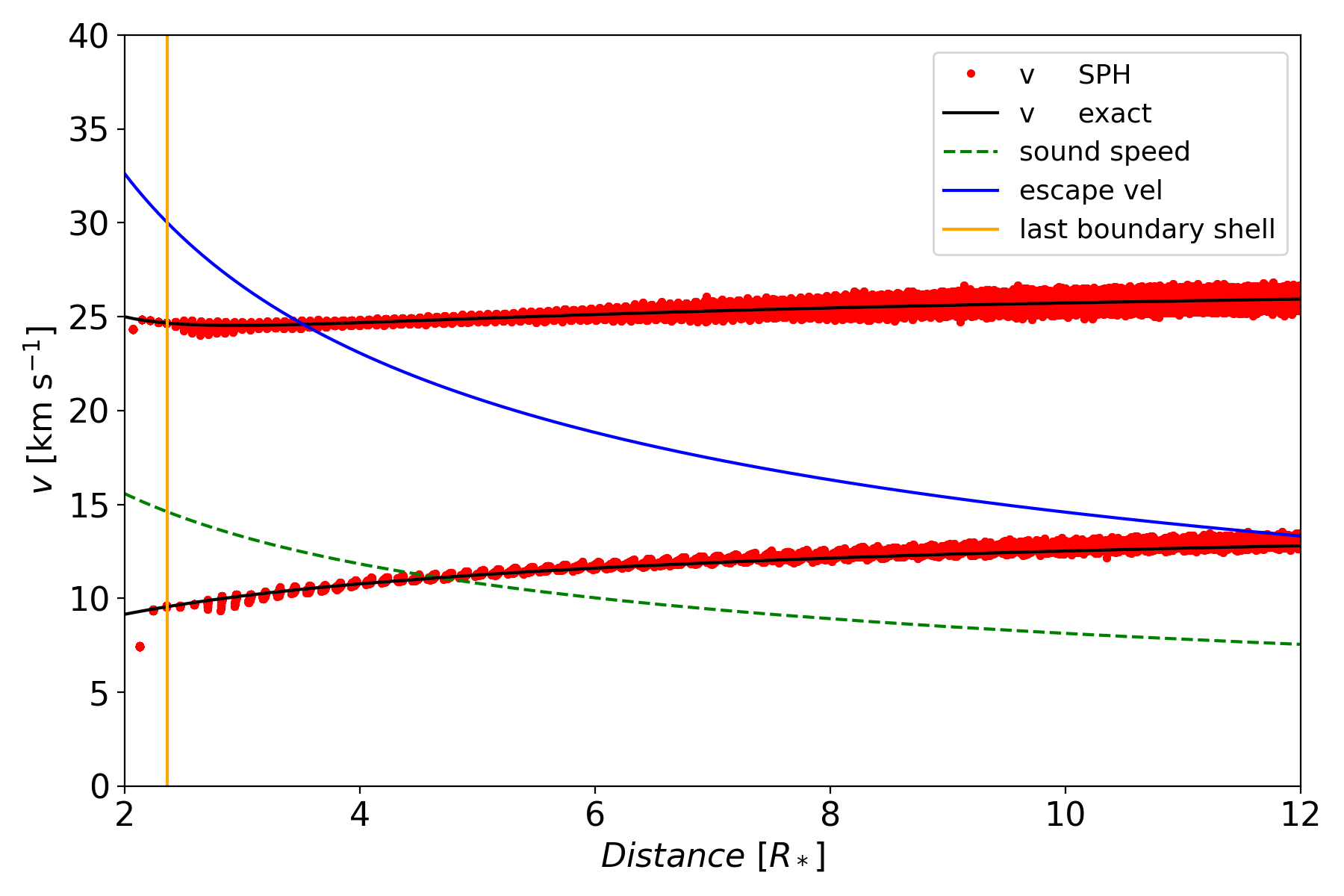}
    \caption{Adiabatic velocity profiles of a transonic (\emph{bottom}) and supersonic (\emph{top}) SPH wind solution, in red. The analytic solution is given by the black curves. }
    \label{fig:adia}
\end{figure}

In this section, we discuss the performance of the implemented wind physics in different idealised thermodynamical contexts. In particular, we investigate the ability of the code to predict transonic and supersonic wind dynamics. Consequently, we focus the discussion on the velocity profile of the wind.
For all studied cases, the physical parameters are described in Table \ref{modpar1d}. Furthermore, the resolution parameter $q$ is set to 15 (see Eq.~\ref{equi}), and the spacing of the shells satisfies the condition in Eq.~(\ref{wsscond}), unless stated otherwise. In practice, we use $w_{ss} = 1.0$. Finally, to position the sonic radius beyond the radius of the handled shells an initial sound speed  $c_\mathrm{s, iso} = 13.12$\kms{} was assumed (corresponding to $T_\mathrm{iso}=1.25 \times 10^4$~K). In all cases presented here and in the following sections, conditions resulting in a breeze-type solution (see Sect.~\ref{sec:innerwind}) were deliberately avoided because the SPH solver was unable to recover it properly.

Figure \ref{fig:iso} shows the velocity profiles of an isothermal SPH wind solution, in the transonic and the supersonic regimes. Shown in black is the analytic solution as derived in Section \ref{sec:innerwind}. The SPH solution shows good agreement with the analytic Parker wind prediction, exhibiting Gaussian 1$\sigma$ dispersion values of 0.22\kms{} and 0.35\kms{} for the supersonic and transsonic solutions, respectively. However, in the subsonic regime of the transonic solution small oscillations can be seen in the SPH solution when the wind is released by the last boundary shell. We henceforth refer to these oscillations as sonic-point oscillations, which are the consequence of the high sensitivity of the transonic solution to the initial conditions. When the SPH particles are released from the last handled shell, the SPH particles rearrange and disperse, which puts the particles on a distribution of interacting quasi-transonic trajectories. However, these oscillations do not affect the overall behaviour of the SPH gas, since beyond the sonic point they die down and the SPH solution nicely follows the analytic solution.

Similar benchmarks were performed for an adiabatic wind, with a polytropic index $\gamma= 1.4$ and are illustrated in Fig.~\ref{fig:adia}. As for the isothermal case, both the trans- and super-sonic simulations tightly follow the analytic Parker wind solution, exhibiting Gaussian 1$\sigma$ dispersion of 0.16\kms{} and 0.29\kms{} for the supersonic and transsonic cases, respectively. Very weak sonic-point oscillations can be seen for the sub-sonic portion of the trans-sonic simulation.

\begin{figure}
    \centering
    \includegraphics[width=9cm]{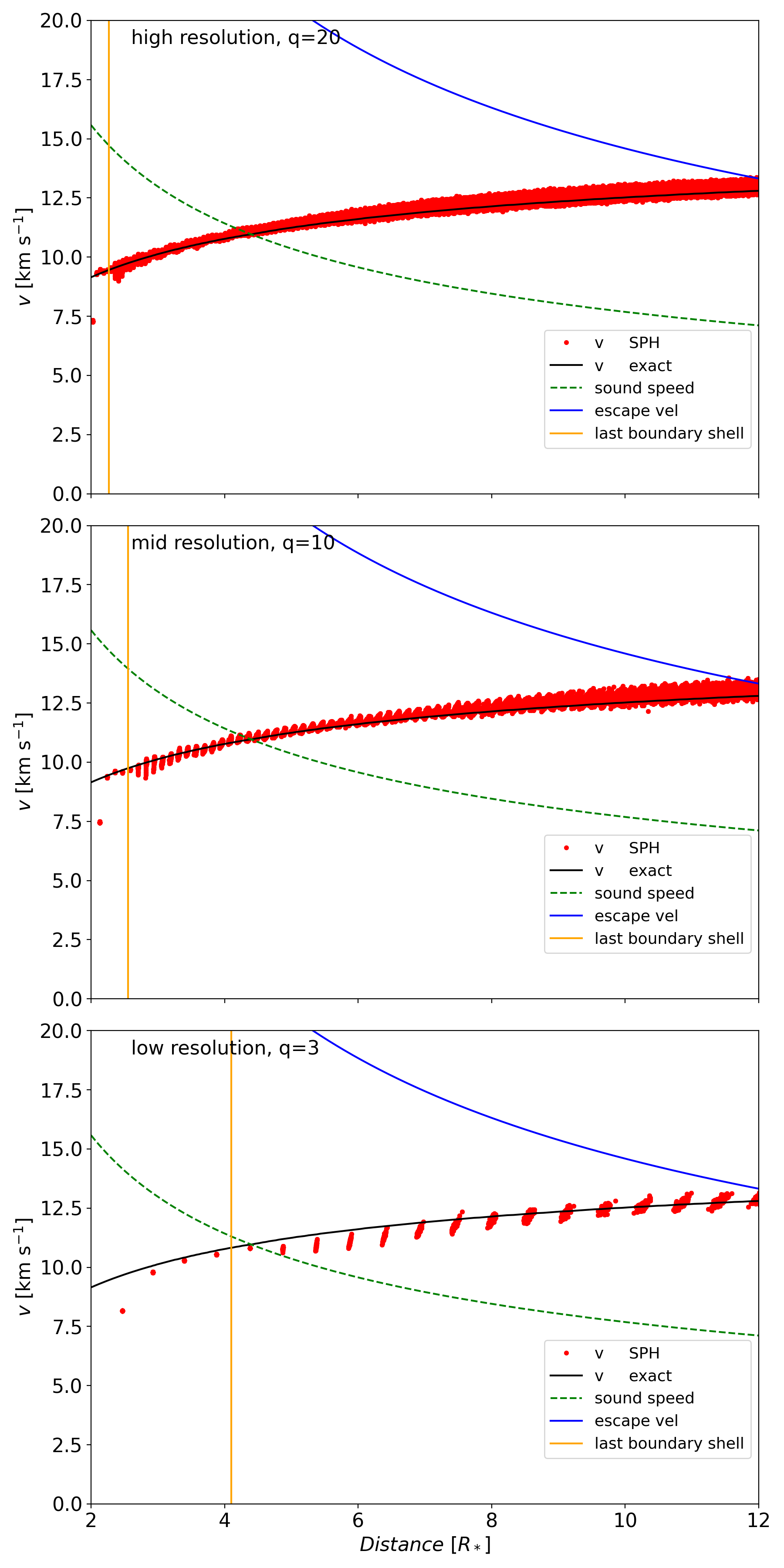}
    \caption{Effect of the resolution setup on the predicted dynamics of an transonic adiabatic wind, for a fixed value of $w_\mathrm{ss}$. The value of $q$ is chosen such that SPH particle mass is decreased by a factor ten from the top to the middle panel, and again from the middle to the bottom panel.}
    \label{resstudy}
\end{figure}

In Fig.~\ref{resstudy}, we show the effect of not properly adjusting the resolution to resolve the critical length-scale of the simulation. We demonstrate the effect by performing a resolution study on the adiabatic transsonic wind case. From the bottom to the top panel, the resolution parameter $q$ is increased from 3 to 10 to 20, which corresponds to a factual decrease of the SPH particle mass with a factor $\sim 10$ at each step. The number of handled shells is 5, which for the simulation with $q=3$ (bottom panel) nearly completely fill up the subsonic region. Consequently, the transonic passage is not properly sampled, and the SPH solution sways around the analytic curve. For the simulations at higher resolution, the transonic passage is nicely resolved, and the SPH solution tightly follows the analytic curve.

\subsection{Effect of Bowen dust opacity prescription}

In this section, we discuss the effect on the SPH wind dynamics of the activation of $\Gamma$ in Eq.~(\ref{motion}). As in the previous section, we focus solely on the velocity profile of the wind.
For all studied cases, the resolution parameter $q$ is set to 10, the number of handled shells is five, and the spacing of the shells satisfies the condition in Eq.~(\ref{wsscond}), unless otherwise stated.

\begin{figure}
    \centering
    \includegraphics[width=9.5cm]{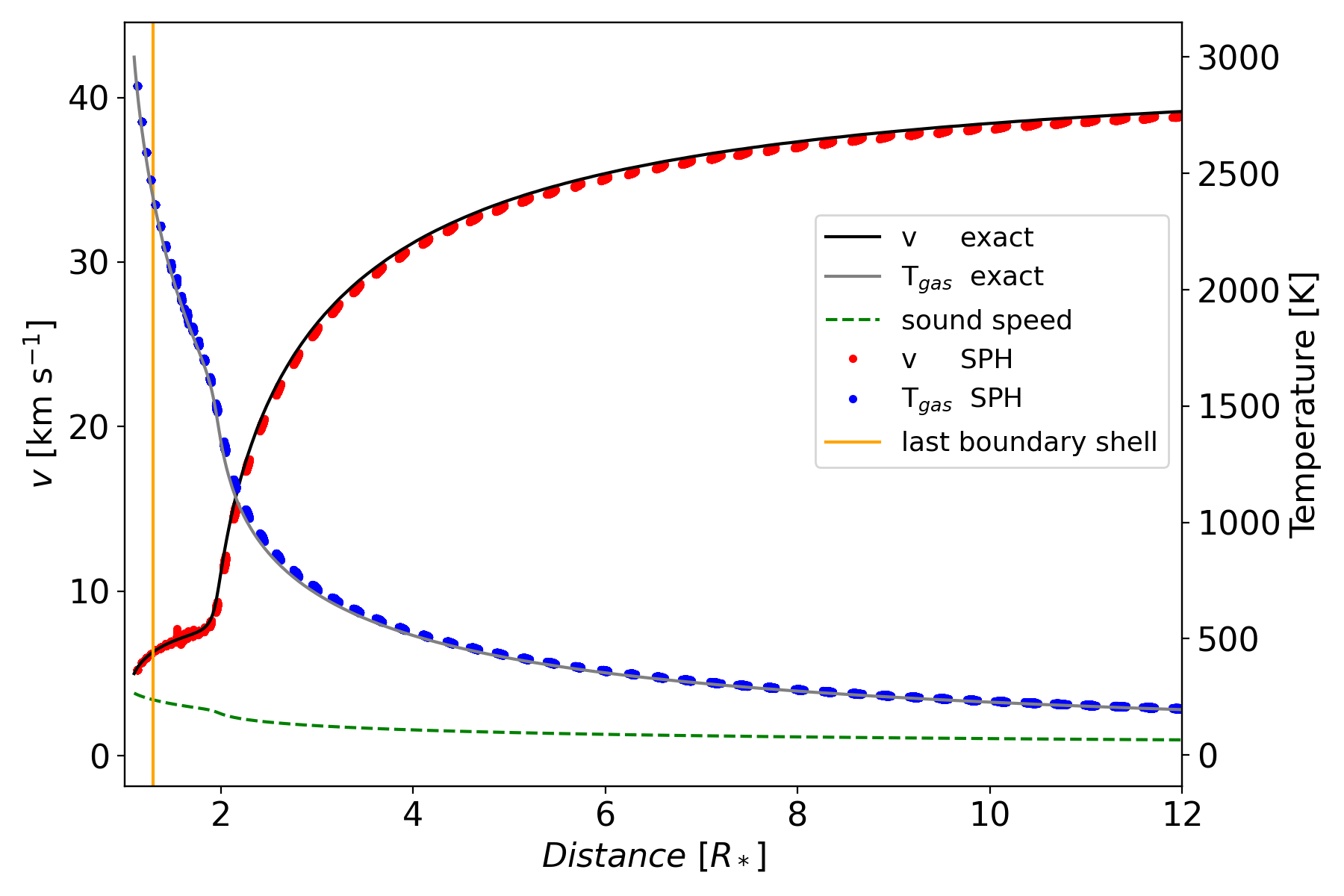}
    \caption{Acceleration profile of a free, steady, polytropic wind with the Bowen dust opacity prescription activated around $\sim 3$~au. The temperature is shown in blue.}
    \label{fig:bowenkappa}
\end{figure}

In his seminal paper, \cite{Bowen_1988} proposed a very simple parametrisation of the dust opacity as a function of temperature. In this formulation, the dust opacity writes
\begin{equation} \label{bowenstep}
\kappa_\mathrm{d}(T_\mathrm{eq}) =  \frac{\kappa_\mathrm{max}}{1+\exp[(T_\mathrm{eq}-T_\mathrm{cond})/\delta]},
\end{equation}
where $\kappa_\mathrm{max}$ is the maximum value the dust opacity can reach, $\delta$ is a measure for the temperature range over which condensation occurs, and $T_\mathrm{cond}$ the dust condensation temperature which depends on the star's composition (C/O ratio). $T_\mathrm{eq}$ is the dust equilibrium temperature which value depends on how the gas and dust are thermally coupled. A correct determination of  $T_\mathrm{eq}$ would require radiative transfer calculations which is outside the scope of this paper (see Sect.~\ref{sec:discussion}) so for the moment we assume that $T_\mathrm{eq}$ is set to the gas temperature.

Though the Bowen parameterization is a severe approximation of the opacity-forming processes, we use this prescription to showcase in Fig.~\ref{fig:bowenkappa} the activation of dust formation on the SPH dynamics. In this setup, a polytropic wind is launched with an initial speed of 5\kms{} from a $1.5 \Msun$ cool star with a surface temperature of 3000~K and a luminosity of $2 \times 10^4 \Lsun$ and with $\mu = 2.38$, corresponding to hydrogen in molecular form. The injection radius is set to 1.1~au. Under these initial conditions, a free wind without pulsations and cooling, and the particles would ballistically fall back onto the stellar surface. These critical ingredients will be the topics of follow-up papers, as discussed in Sect.~\ref{sec:discussion}. To circumvent this problem we introduced an additional  acceleration parameter  $\alpha$ in Eqs.~(\ref{motion}) and (\ref{1d2}) to the Eddington factor $\Gamma$. This parameter $\alpha$, takes on a value between zero and one, and serves solely to counteract the gravitational potential. Although this parameter is used in an ad-hoc manner in our context, it could in principle be related to the energy injected by the pulsations and the propagation of a pressure wave \citep[e.g.][]{Struck2004,Mattsson2016}. This technique is commonly used in hydrodynamical wind simulations that do not calculate dust acceleration \citep[e.g.][]{mastrodemos_morris_1998,Kim2012,Liu2017}. Consequently, the gravity term in Eqs.~(\ref{motion}) and (\ref{1d2}) now reads
\begin{equation} \label{gravamend}
- \frac{G M_*}{r^2} (1 - \alpha - \Gamma),
\end{equation}
where the meaning of all other parameters has remained unchanged.

The profile in Fig.~\ref{fig:bowenkappa} was generated by setting $\alpha$ equal to unity and an initial velocity at the injection radius of $v_\mathrm{inj} = 5$\kms. We note that setting $\alpha=1$ will effectively overestimates the true acceleration of the wind. This is of no consequence for this exercise, since the aim is to illustrate the mechanism of dust acceleration in action. We set the dust condensation temperature to 1500~K, and the width of the condensation temperature range $\delta$ to 60~K and the polytropic index of the gas to $\gamma=1.4$.
Finally a value of $\kappa_\mathrm{max}=1.2~\mathrm{cm^2\,g^{-1}}$ is chosen, yielding $\Gamma(T_\mathrm{eq}=T_\mathrm{cond}) \simeq 0.6$ (using Eqs.~\ref{dustGamma} and \ref{bowenstep}).

The SPH gas (red in Fig.~\ref{fig:bowenkappa}) undergoes a double acceleration. The first stage, which occurs between 1 and $\sim 2$~au, is a consequence of the inner wind pressure gradient caused by the stellar temperature boundary condition. Thanks to the parameter $\alpha$ set to unity in Eq.~(\ref{gravamend}), the gas can freely escape and accelerate outwards. At a radius of $\sim 2$~au, the temperature of the gas drops below the condensation temperature  which activates the dust opacity and raises $\Gamma$ in Eq.~(\ref{dustGamma}). This results in a second acceleration that brings the velocity of the gas around 37\kms.
The dynamics of the SPH particles nicely follows the analytic curve. Especially in the first acceleration stage, the analytic curve (black) is centered on the SPH solution. When dust starts forming and triggers the secondary acceleration, the velocity of the SPH is slightly underpredicted with respect to the analytical solution. It appears that when the radiative acceleration is reduced, for example when the value of $k_\mathrm{max}$ in Eq.~(\ref{bowenstep}) is lowered, the agreement with the theoretical line is even better. This effect is likely related to the accuracy of the time integration scheme which is 2nd order accurate in SPH compared to the 4th order Runge-Kutta scheme used for the 1D solution. But the overall agreement is within a few percent.
As a final note, we remark that the low surface temperature of the star implies that the SPH particles are launched fully supersonically. Consequently, the instabilities associated with passing the sonic point do not arise in this simulation (see Figs.~\ref{fig:iso} and \ref{fig:adia}).

\subsection{Steady wind with dust nucleation}

\begin{figure}
    \centering
\includegraphics[width=9cm]{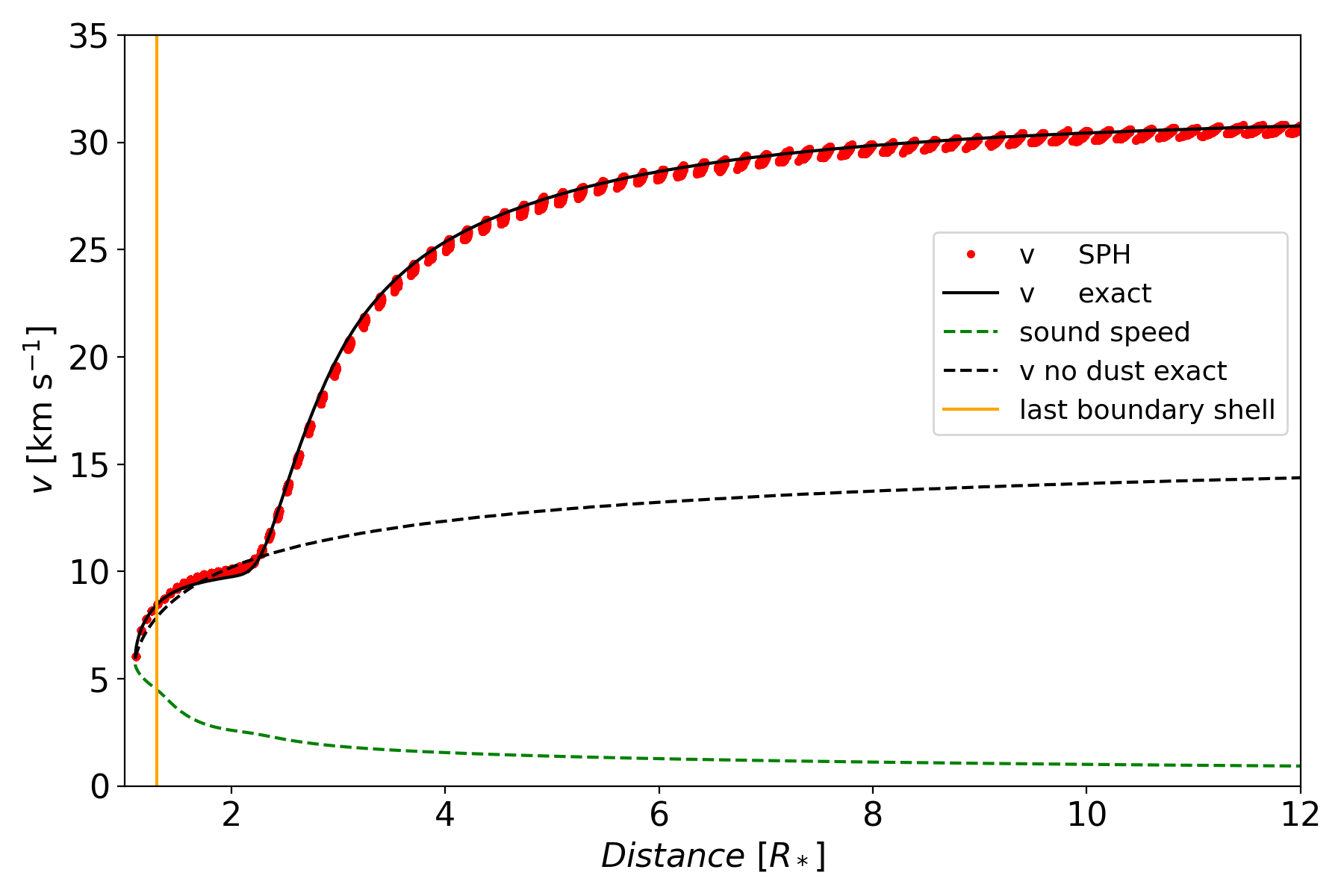}
    \caption{Acceleration profile of a free wind for which the dust opacity is calculated using the nucleation formalism.}
    \label{fig:nucleation}
\end{figure}

\begin{figure*}
    \centering
    \includegraphics[width=8.5cm]{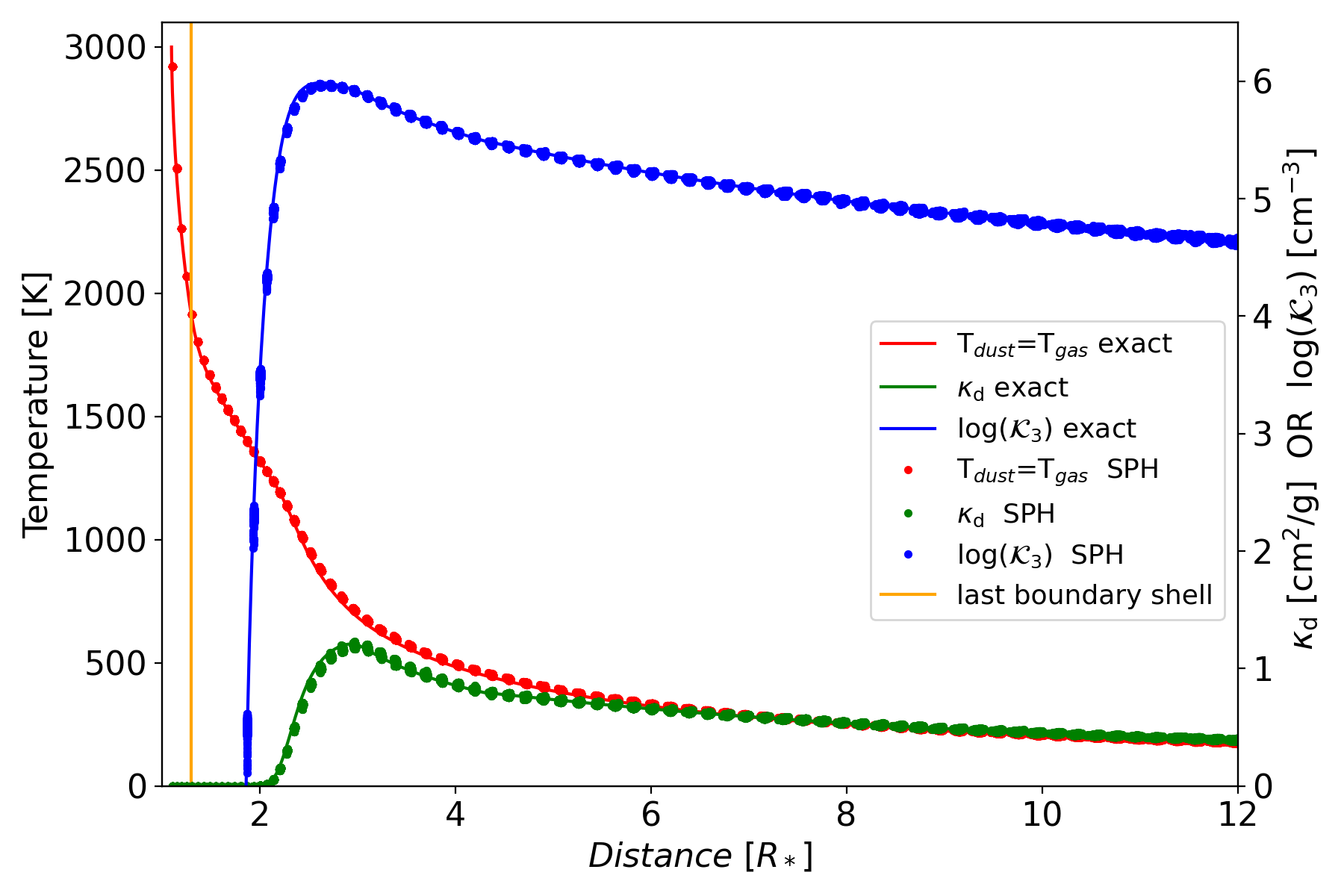}
    \includegraphics[width=8.5cm]{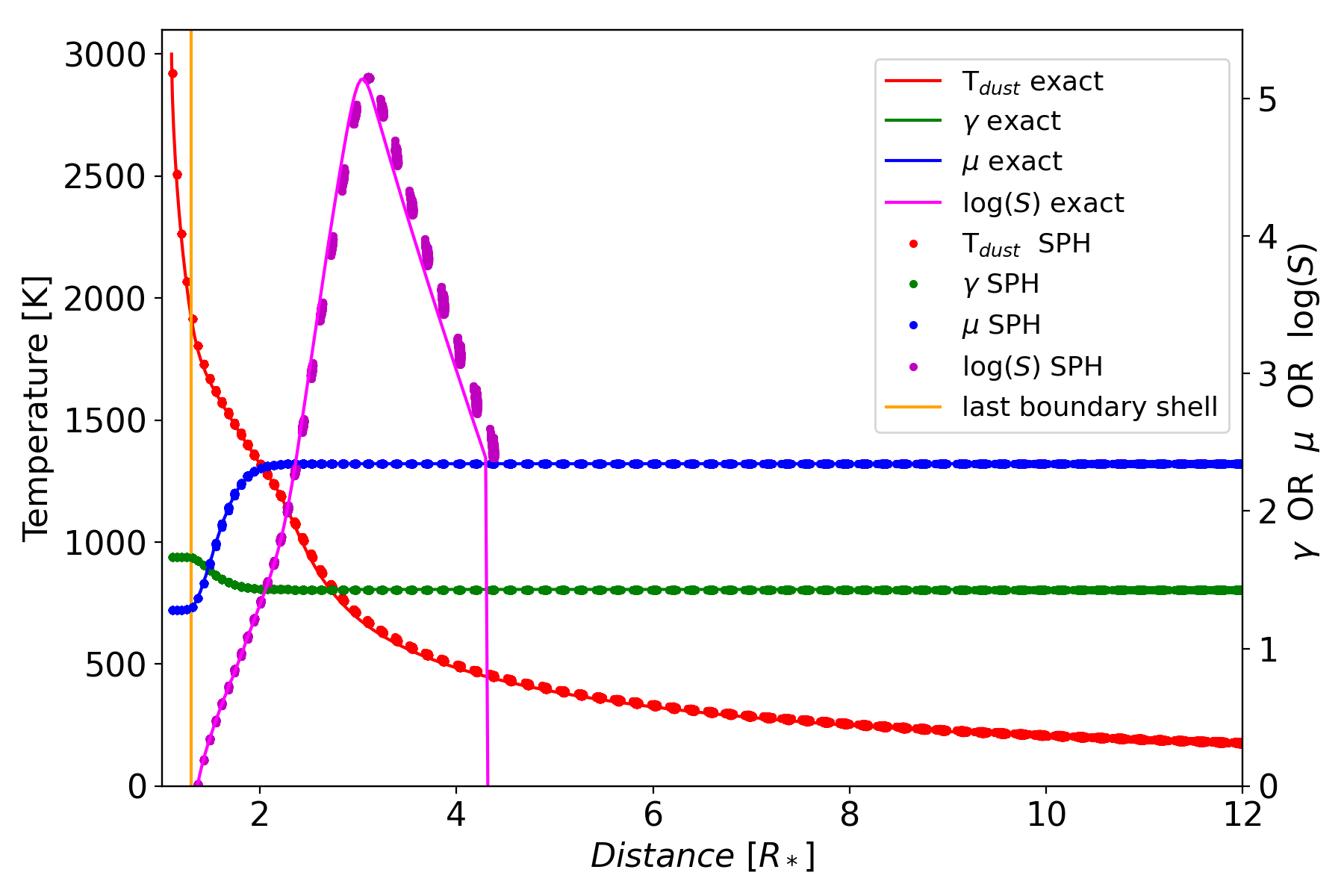}
    \caption{Radial profiles of chemical and dust-related quantities. The temperature of the SPH gas, with which the dust is assumed to be fully thermalised, is shown in red.
    \emph{Left:} The logarithm of the third moment $\mathcal{K}_3$ (blue), which represents the number density of condensing monomers from the gas phase into the dust, shown together with and the dust opacity profile of the medium as the dust condensation temperature threshold is crossed (green). \emph{Right:} The mean molecular weight $\mu$ (blue), polytropic index $\gamma$ (green), and the base 10 logarithm of the supersaturation ratio $S$ (magenta) of the gas in the wind. A supersaturation ratio above unity implies favourable conditions for dust formation.}
    \label{fig:kappa}
\end{figure*}

In Fig.~\ref{fig:nucleation}, we show the velocity profile of a free ($\alpha=1$), steady wind that includes the full computation of the dust-nucleation described in Sect.~\ref{sec:physics}. The numerical setup is nearly identical to the one used with the Bowen dust-opacity prescription described above, except that now the composition of the gas, and consequently its mean molecular weight $\mu$ and polytropic index $\gamma$, are set by the solution of the chemical network described in Sect.~\ref{sec:molabund}. The chemistry is set to a carbon to oxygen ratio of two. And contrarily to the previous case studies, the velocity and temperature profiles now depend on the value of the mass loss rate because the chemical and nucleation processes are explicit functions of the density. Indeed, the independence of the stationary wind velocity profile on $\dot{M}$ comes from the fact that in the absence of cooling
and chemistry, the density $\rho$ acts only as a scaling factor in Eqs.~(\ref{momconcool}) and (\ref{tempcool}). Integration of these equations yield rho-invariant functions $v(r)$ and $T(r)$, with the density profile then deduced from Eq.~(\ref{dens1d}) that directly scales with the mass loss rate.
For the simulation showcased in Fig.~\ref{fig:nucleation}, a mass-loss rate of $10^{-5}$\myr, and an initial velocity of $v_\mathrm{inj} = 6$\kms{} was assumed. This initial velocity differs slightly from the Bowen-type dust-opacity-setup in the previous section to avoid the breeze-type solution.

The effect of including dust nucleation on the wind dynamics is significant, and can be readily assessed by comparing the SPH solution (red) and the dust-free profile (black dashed) in Fig.~\ref{fig:nucleation}. While the SPH solution reaches a terminal velocity of nearly 30\kms, its dust-free counterpart (accelerated only via the pressure-gradient at the stellar boundary and $\alpha=1$) barely reaches a speed of 12\kms. The dusty wind also exhibits a double-staged acceleration. The inner-most velocity profile of the dust-free curve matches the SPH solution, indicating that the first acceleration stage is due to the inner boundary conditions\, similar to what was found in the previous section (see Fig.~\ref{fig:bowenkappa}). However, around 3~$R_*$ a much steeper acceleration of the SPH gas can be seen, which is not reproduced by the dust-free curve, and is therefore solely associated with dust formation.

The formation of dust can be quantified either by the rate of condensation from the gas-phase, of by their effect on the dust opacity. Both these quantities are shown in the left panel of Fig.~\ref{fig:kappa}. The sudden increase in the nucleation rate $\mathcal{K}_3$ (blue) from about zero to $\sim 10 ^{6}$~cm$^{-3}$  around $\sim 3~R_*$ coincides with a dramatic increase in dust opacity (green), which explains the sudden acceleration of the SPH particles. At $\sim 4~R_*$, the nucleation rate reaches a plateau, and the corresponding dust opacity decreases as a consequence of decreasing density.

The supersaturation ratio $S$, shown in purple in the right panel of Fig.~\ref{fig:kappa}, can also be used as a proxy for dust formation. For values above unity, the partial pressure of gas-phase carbon exceeds that of solid-state carbon, indicating favourable condensation conditions. Indeed, the sudden increase in relative condensation rate $\mathcal{K}_3$ seen in the left panel of Fig.~\ref{fig:kappa} occurs as $S$ becomes much larger than unity. At distances slightly larger than 4~au the temperature reaches 500~K, below which the simplified chemistry is activated (all H and C atoms locked in molecules) producing the artificial discontinuity in $S$ (Sect.~\ref{3dhydro}).

The radial decrease in gas temperature causes the hydrogen in the wind to transition from being fully monatomic to fully fully diatomic (see also Fig.~\ref{fig:abund}). This is reflected by the rapid change in mean molecular weight $\mu$ and polytropic index $\gamma$ of the gas, seen in the right panel of Fig.~\ref{fig:kappa}. These quantities directly affect the temperature of the gas. Given that these simulations do not include cooling, except for adiabatic compression and expansion, this results in a small bump in the radial temperature profile.

\section{Discussion and future prospects}
\label{sec:discussion}

The simulations performed in Sect.~\ref{sec:3D} show that our implementation of the wind-injection boundary, as well as the calculations involving dust opacity can be used to model cool and dusty stellar winds. However, to level-up the modelling of dust-driven winds to match the state-of-the-art knowledge on the physical mechanisms governing these outflows, many additional physical and chemical ingredients have yet to be accounted for \citep[a comprehensive summary of theoretical challenges can be found in][]{Haworth2016}.

The stellar surface pulsations, driven by convective motions and an essential ingredient for launching a dusty wind, have been circumvented in our approach by enforcing a ``free'' wind by setting $\alpha=1$. This is a common assumption in the modelling of 3D winds \citep[e.g.][]{Jahanara2005,Kim2012,Liu2017} but when dust formation is involved a more correct treatment of the wind launching mechanism is needed \citep{Freytag2017}. We will adopt a dynamical model as originally done by \citet{Bowen_1988}, where the pulsations are simulated by a radially oscillating inner boundary, which can be considered to act as a piston located at the base of the photosphere. A recent implementation of this approach in the SPH context is described in \cite{Aydi2022}. Considering the modelling of large-scale convection as the source of pulsation as done in \cite{Freytag2017} would require the implementation of an energy transport equation in \phantom{} and is beyond the scope of our ambitions for this project.

Another crucial issue is related to the chemistry and cooling processes \citep{Boulangier2018}. The use of a polytropic index does not provide an accurate estimate of the thermodynamics of the gas because several processes such as radiative cooling by line transitions, chemical, and continuum cooling, will significantly modify the internal energy.
In turn, this will impact the dust formation process and consequently the wind acceleration. In the context of wind-companion interactions, the absence of cooling has been shown to prevent the formation of an accretion disk around an AGB companion \citep{theuns_jorissen_1993}, with a dramatic impact on the derived accretion and angular momentum transfer rates and potentially on the secular evolution of the binary system. Furthermore, it will impact the flow morphology, affecting the interpretation of key features in the observations.

Related to this issue, is the correct treatment of the radiative acceleration which is not considered as we assume an optically thin wind. However, if shocks form in the outflow, the opacity can locally increase dramatically and as we showed provide an extra acceleration which will impact the wind morphology. The treatment of radiative transfer is a very challenging and computationally intensive endeavour but is not beyond reach.
Different approaches has been explored to solve the radiative transfer equation and are compared in the context of cosmological simulations, in \cite{Iliev2006,Iliev2009}. They include ray tracing algorithms where each source casts a number of rays and the equation of radiative transfer (RT) is integrated along the ray \citep[e.g.][]{Kessel2000,Susa2006,Altay2013,Chen2020}. If scattering is neglected, this method reduces to calculating the optical depth along the ray. Another option is to solve the angular moments of the radiative transfer equation using specific closure relations and/or flux limiters \citep[e.g.][]{Gnedin2001,Whitehouse2005,Aubert2008,Petkova2009,Skinner2013,Chan2021}. Other methods involving transport on unstructured meshes such as Delaunay tessellations \citep[e.g.]{Rijkhorst2006,Pawlik2008,Petkova2011,Petkova2021} have also been proposed. A final  option is  to delegate the resolution of the RT problem to an external code. This has been done in \phantom{} in dusty proto-planetary disk simulations where the output of the SPH simulation were regularly processed by the RT code MCFOST code \citep[e.g.][]{Pinte2019} providing updated values for the gas and dust temperatures of each particles.
Considering that we are mostly interested in ``simple'' systems consisting of one or two sources, the ray tracing approach seems to be a promising avenue. We will explore this possibility in a the future following an approach similar to \cite{Hasegawa2010}.

A great simplification of the presented model is the absence of dust-gas drift. This process is important because conceptually the absence of drift is inconsistent with the dust-driven wind mechanism, which says that the gas is dragged along by the faster, radiatively accelerated dust particles \cite[see e.g. discussion in Sect 4.1 of][]{Mattsson2021}. In addition, as shown by \cite{Kruger1994,Sandin2004,Sandin2008,Sandin2020}, the effect of drift can potentially lead to an increase in the dust production and, in some cases, to a decrease of the mass-loss rates and wind velocities. Fortunately, \phantom{} allows the modelling of dust-gas mixtures in either the standard two-fluid approach where each species is regarded as a separate fluid \citep{Laibe2012} or in the so-called one-fluid approximation \citep{laibe2014a,laibe2014b,Price2015,Hutchison2018} where the evolution of the gas and dust velocities require the tracking of only one additional quantity, the mass-fraction of the dust-grain mixture. This ``diffusion approximation for dust'' holds as long as the stopping time-step is small compared to the hydrodynamical timestep which should be the case for AGB winds in a steady regime outside shocked regions \cite[see Sect 6.8.3 of][]{Gail_Sedlmayr_2014}. Using this particular prescription overcomes some of the limitation associated with the two-fluid approach, and allows the use of much larger timesteps, which speed up the calculations. We are currently working on including this formalism in our dusty wind simulations.

The road to reach a more consistent picture is still long and will proceed step-wise, but even with some missing physics, our simulations \citep{Maes2021,Malfait2021} can already help understand wind morphologies around AGB stars.

\section{Summary}
\label{sec:summary}

In this paper, we present the implementation of a dusty wind into the smoothed particle hydrodynamics code \phantom. The SPH particles are initially distributed on a geodesic sphere and their properties determined by solving the 1D stationary wind equations. To counteract the buildup of a positive pressure gradient at the inner boundary due to the presence of a sink particle at the center, we force a number of fixed boundary shells, whose properties are imposed by the 1D wind equations.

Using this set-up, we have calculated a number of different benchmarking simulations that test the ability of the code to recover the stationary Parker-wind solutions. We show that the SPH solution tightly follows the expected supersonic and transonic wind trajectories. For the transonic case, the SPH solution shows weak oscillations in the subsonic portion of the trajectory. This is a consequence of the fact that the transonic trajectory is very sensitive to the precise initial conditions of the launched particles. The mutual interaction between the SPH particles makes them deviate slightly from this very narrow path, leading to oscillations in their velocity. Nevertheless, when the sonic point is eventually crossed oscillations die down leading to a smooth SPH solution.

The dust component of the wind is implemented according to two different schemes. The first is via an analytical expression that describes how the dust opacity increases once a dust condensation temperature threshold is crossed, as described by \citet{Bowen_1988}. In the second scheme dust nucleation is calculated from the gas composition of the outflow. This composition is obtained by solving a compact carbon-rich chemical network assuming local-thermodynamical-equilibrium. This leads to the formation carbonaceous monomers, that nucleate into dust particles. The dust formation formalism follows the moment equations prescription described by \citet{Gail_Sedlmayr_2014}. This provides a quick way to estimate the rate at which the monomers condense into the dust grains, which can be directly related to the dust opacity.

Assuming a free ($\alpha=1$) optically thin wind, in which the gas and the dust are thermally and mechanically coupled ($\Tg = \Td$ and with no drift velocity), both schemes lead to the establishment of an significant additional opacity term that dramatically impacts the wind dynamics by providing an additional outward acceleration. For the nucleation models, we show how the partial pressures of the carbon-compounds affect the supersaturation ratio, how this directly impacts the rate at which the carbonaceous monomers condense into the dust, and how this further coincides with the vast opacity increase that accelerates the wind.

As such, we illustrate that our implementation of the SPH particle injection set-up, and the nucleation and dust growth prescriptions in the \phantom{} code can be used to successfully model dusty stellar outflows.

\section*{Acknowledgments}
The authors are in debt to H.-P. Gail and C. Dreyer who provided us a long time ago the source of their code that were extremely useful to check and validate our dust routines. We also thank the anonymous referee for very constructive and encouraging comments. L.S. is senior FRS-F.N.R.S. research associate. W.H. acknowledge support from a FRS-F.N.R.S grant. D.J.P. is grateful for Australian Research Council funding via DP180104235 and FT130100034.

\bibliographystyle{aa}
\bibliography{sph_wind}

\appendix

\section{Chemical network}
\label{sec:chemistry}

\begin{figure}[b]
    \centering
    \includegraphics[width=\columnwidth]{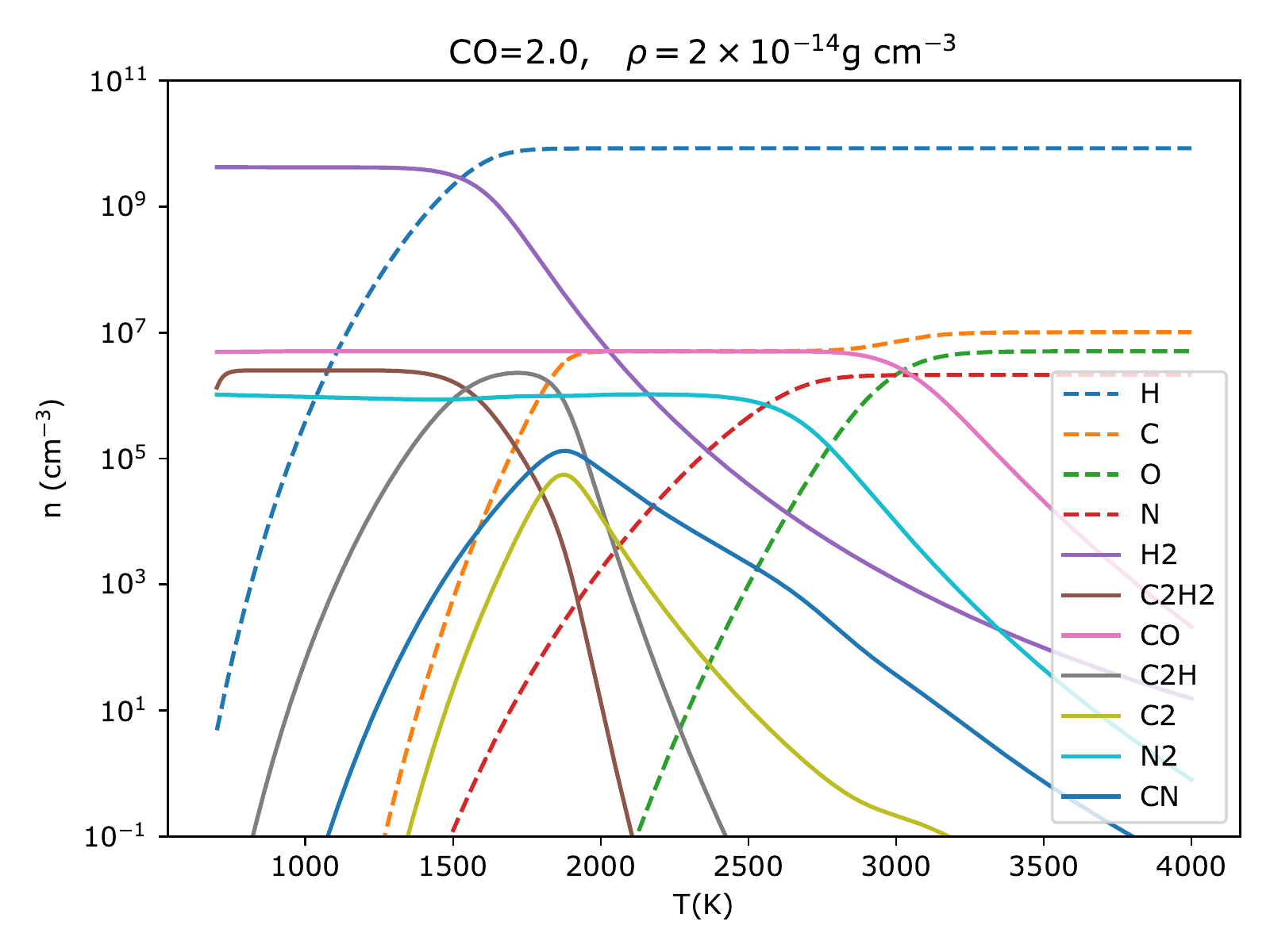}
    \caption{Evolution of the number density of the most abundant species as a function of temperature for the envelope composition given in Table~\ref{tab:abundances} and for a density $\rho = 2\times 10^{-14} \, \mathrm{g~cm^{-3}}$ and a carbon to oxygen ratio C/O=2.}
    \label{fig:abund}
\end{figure}

The chemical network used to compute the carbon abundance needed to estimate the opacity includes the following 26 reactions :
\begin{eqnarray*}
\mathrm{H}  + \mathrm{H} & \rightarrow &  \mathrm{H}_2  \\
\mathrm{O}  + \mathrm{H} & \rightarrow &  \mathrm{OH}  \\
\mathrm{O} + 2\mathrm{H} & \rightarrow &  \mathrm{H}_2\mathrm{O}  \\
\mathrm{C}  + \mathrm{C} & \rightarrow &  \mathrm{C}_2  \\
\mathrm{C}  + \mathrm{O} & \rightarrow &  \mathrm{CO}  \\
\mathrm{C}  + 2\mathrm{O} & \rightarrow &  \mathrm{CO}_2  \\
\mathrm{C}  + 4\mathrm{H} & \rightarrow &  \mathrm{CH}_4  \\
\mathrm{C}  + \mathrm{H} & \rightarrow &  \mathrm{C}_2\mathrm{H}  \\
\mathrm{H}  + \mathrm{H} & \rightarrow &  \mathrm{C}_2\mathrm{H}_2  \\
\mathrm{N}  + \mathrm{N} & \rightarrow &  \mathrm{N}_2  \\
\mathrm{N}  + 3\mathrm{H} & \rightarrow &  \mathrm{NH}_3  \\
\mathrm{C}  + \mathrm{N} & \rightarrow &  \mathrm{CN}  \\
\mathrm{CN} + \mathrm{H} & \rightarrow &  \mathrm{HCN}  \\
\mathrm{Si} + \mathrm{Si} & \rightarrow &  \mathrm{Si}_2  \\
\mathrm{Si} + 2\mathrm{Si} & \rightarrow &  \mathrm{Si}_3  \\
\mathrm{Si} + \mathrm{O} & \rightarrow &  \mathrm{SiO}  \\
2\mathrm{Si} + \mathrm{C} & \rightarrow &  \mathrm{Si}_2\mathrm{C}  \\
\mathrm{Si} + 4\mathrm{H} & \rightarrow &  \mathrm{SiH}_4  \\
\mathrm{S}  + \mathrm{S} & \rightarrow &  \mathrm{S}_2  \\
\mathrm{S}  + \mathrm{H} & \rightarrow &  \mathrm{HS}  \\
\mathrm{S}  + 2\mathrm{H} & \rightarrow &  \mathrm{H}_2\mathrm{S}  \\
\mathrm{Si} + \mathrm{S} & \rightarrow &  \mathrm{SiS}  \\
\mathrm{Si} + \mathrm{H} & \rightarrow &  \mathrm{SiH}  \\
\mathrm{Ti} + \mathrm{O} & \rightarrow &  \mathrm{TiO}  \\
\mathrm{Ti} + 2\mathrm{O} & \rightarrow &  \mathrm{TiO}_2 \\
\mathrm{Ti} + \mathrm{S} & \rightarrow &  \mathrm{TiS}
\end{eqnarray*}
The dissociation constants for the different reactions are calculating using Eqs.~(\ref{dissh2}) and (\ref{gibbsen}) where the Gibbs energies are taken from the fits of the JANAF tables \citep{Chase_etal_1986}. They are expressed in terms of polynomials of the temperature $T$ of the form
\begin{equation}
    \Delta G_\mathrm{f} = \frac{a}{T}+b+cT+dT^2+eT^3,
\end{equation}
where the coefficients $a,\ b,\ c,\ d$ and $e$ can be found in \cite{sharp1990}.
For TiS, the fit is taken from \cite{Tsuji1973}.

Figure \ref{fig:abund} illustrates how the abundance of the main species vary as a function of temperature. At low temperatures ($T \la 1400$~K) all hydrogen is in the form of H$_2$ and the second most abundant species is CO. As the temperature increases, H atoms become available and C$_2$H$_2$ dissociates. At higher temperature ($T > 2500$~K) all the material returns to an atomic state. We note that CO is the most resistant molecule and at this selected density it can survive up to $T\approx 3000$~K.

\section{Computing the net growth rate}
\label{detnet}
With the four chemical species C, C$_2$, C$_2$H, and C$_2$H$_2$ that we consider for the
evolution of carbon grains, the net growth rate (Eq.~\ref{taurate}) writes
\begin{equation}
  \frac{1}{\tau} = \frac{1}{\tau_\mathrm{C}} + \frac{1}{\tau_{\mathrm{C}_2}} +
  \frac{1}{\tau_{\mathrm{C}_2\mathrm{H}}} + \frac{1}{\tau_{\mathrm{C}_2\mathrm{H}_2}}.
\end{equation}
The first two terms of Eq.~(\ref{taurate}), which include $P_\mathrm{C}$ and
$P_{\mathrm{C}_2}$, account for homogeneous grain growth and evaporation. In chemical
equilibrium, the general form of the second term reads \citep{Gauger_etal_1990},
\begin{equation}
  \frac{1}{\tau_{\mathrm{C}_2}} = \frac{2 \A_1 \alpha(2) P_{\mathrm{C}_2}}{\sqrt{2 \pi k \Tg\,
      m_{\mathrm{C}_2}}} \left[1 - \frac{\alpha_*(2)}{\Sat^2 b_{\mathrm{C}_2}}\right],
\end{equation}
where $\alpha(2)$ is an averaged sticking efficiency for a dimer, in this case for C$_2$,
and the factor $\alpha_*(2)$ accounts for non-thermal equilibrium (non-TE) effects,
\begin{equation}
    \alpha_*(2) = \frac{\mathring{\alpha}(2)}{\alpha(2)}.
\end{equation}
The quantity $\mathring{\alpha}$ denotes the averaged thermal equilibrium (TE) reaction efficiency when the gas temperature equals the grain temperature. We assume TE for the reaction efficiency and use the sticking coefficient $\alpha_2$ (see sect.~\ref{netrate}) both for $\mathring{\alpha}(2)$ and $\alpha(2)$. Similarly, we use $\alpha_1$ in the expression of
$\tau_\mathrm{C}^{-1}$. In chemical equilibrium, and when the gas has the same temperature as the dust, the correction factors $b$ read $b_\mathrm{C} = b_{\mathrm{C}_2} = 1$.

The third and fourth terms of Eq.~(\ref{taurate}) that include the contributions from the molecules C$_2$H and C$_2$H$_2$, account for heteromolecular growth and chemical sputtering. The general form of these terms has been presented by \citet{Gauger_etal_1990}, and for the third term it reads
\begin{equation}
    \frac{1}{\tau_{\mathrm{C}_2\mathrm{H}}} = \frac{2 \A_1
        \alpha_{\mathrm{C}_2\mathrm{H}}^\mathrm{c}(2)
        P_{\mathrm{C}_2\mathrm{H}}}{\sqrt{2 \pi k \Tg\,
        m_{\mathrm{C}_2\mathrm{H}}}} \left[1 - \frac{\alpha_*^\mathrm{c}(2,
        \mathrm{C}_2\mathrm{H})}{\Sat^2 b_{\mathrm{C}_2\mathrm{H}}^\mathrm{c}}\right],
\end{equation}
where the factor $\alpha_*^\mathrm{c}(2, \mathrm{C}_2\mathrm{H})$ expresses non-TE effects,
\begin{equation}
    \alpha_*^\mathrm{c}(2, \mathrm{C}_2\mathrm{H}) =
        \frac{\mathring{\alpha}_{\mathrm{C}_2\mathrm{H}}^\mathrm{c}(2)
        \beta_{\mathrm{C}_2\mathrm{H}}^\mathrm{c}(2)}
        {\alpha_{\mathrm{C}_2\mathrm{H}}^\mathrm{c}(2)
        \mathring{\beta}_{\mathrm{C}_2\mathrm{H}}^\mathrm{c}(2)}.
\end{equation}
$\alpha$ denotes the reaction efficiency of the growth reaction, and
$\beta$ the efficiency of the reverse reaction, $\mathring{\beta}$ being
the TE value of the latter. We assume that both are equal ($\alpha^{c} =
\beta^{c}$) and set $\alpha_*^\mathrm{c}(2, \mathrm{C}_2\mathrm{H}) =
\alpha_*^\mathrm{c}(2, \mathrm{C}_2\mathrm{H}_2) = 1$. Moreover, we set
$\alpha_{\mathrm{C}_2\mathrm{H}}^\mathrm{c}(2) =
\alpha_{\mathrm{C}_2\mathrm{H}_2}^\mathrm{c}(2) = \alpha_2$.

\end{document}